\documentclass[a4paper,11pt]{article}
\usepackage{jcappub}
\usepackage{hyperref}
\usepackage{orcidlink}

\usepackage{amsmath}
\usepackage{amssymb}

\usepackage[utf8]{inputenc}
\usepackage[a4paper, left=2.5cm, right=2.5cm, bottom=2.5cm, top=2.5cm]{geometry}
\usepackage{comment}
\usepackage{graphicx}
\usepackage{bbold}
\usepackage[normalem]{ulem}
\graphicspath{ {figures/} }

\newcommand{\altmtl}{altMTL}
\newcommand\aver[1]{\left\langle#1\right\rangle}
\newcommand\abs[1]{\left|#1\right|}
\newcommand\vc[1]{\mathbf{#1}}
\newcommand\Leg[1]{\mathcal{L}_{#1}}
\newcommand\vx{\vc{x}}
\newcommand\vd{\vc{d}}

\newcommand\vt{\vc{t}}

\newcommand\vr{\vc{r}}
\newcommand\vs{\vc{s}}

\newcommand\vk{\vc{k}}

\newcommand\vmo{\vc{m}_{\rm o}}
\newcommand\vmt{\vc{m}_{\rm t}}
\newcommand{\kt}{k\textsubscript{\rm t}}
\newcommand{\ko}{k\textsubscript{\rm o}}

\newcommand\cut{\rm cut}
\newcommand{\Mpc}{\rm Mpc}
\newcommand{\Gpc}{\rm Gpc}
\newcommand{\Pt}{P\textsubscript{\rm t}}
\newcommand{\Po}{P\textsubscript{\rm o}}

\title{Mitigation of DESI fiber assignment incompleteness effect on two-point clustering with small angular scale truncated estimators}

\emailAdd{mathilde.pinon@cea.fr}
\emailAdd{arnaud.de-mattia.cea.fr}
\emailAdd{etienne.burtin@cea.fr}

\affiliation{Affiliations are in Appendix \ref{sec:affiliations}}

\author[1]{{M.~Pinon}\orcidlink{0009-0009-3228-7126},}
\author[1]{{A.~de~Mattia}\orcidlink{0000-0003-0920-2947},}
\author[2]{{P.~McDonald}\orcidlink{0000-0001-8346-8394},}
\author[1]{{E.~Burtin},}
\author[1]{{V.~Ruhlmann-Kleider}\orcidlink{0009-0000-6063-6121},}
\author[3,4]{{M.~White}\orcidlink{0000-0001-9912-5070},}
\author[5]{{D.~Bianchi}\orcidlink{0000-0001-9712-0006},}
\author[6,7,8]{{A.~J.~Ross}\orcidlink{0000-0002-7522-9083},}
\author[2]{{J.~Aguilar},}
\author[9]{{S.~Ahlen}\orcidlink{0000-0001-6098-7247},}
\author[10]{{D.~Brooks},}
\author[2]{{R.~N.~Cahn}\orcidlink{0000-0003-2748-0641},}
\author[2]{{E.~Chaussidon}\orcidlink{0000-0001-8996-4874},}
\author[2]{{T.~Claybaugh},}
\author[11]{{S.~Cole}\orcidlink{0000-0002-5954-7903},}
\author[12]{{A.~de la Macorra}\orcidlink{0000-0002-1769-1640},}
\author[13]{{B.~Dey}\orcidlink{0000-0002-5665-7912},}
\author[10]{{P.~Doel},}
\author[14,15]{{K.~Fanning}\orcidlink{0000-0003-2371-3356},}
\author[16,17]{{J.~E.~Forero-Romero}\orcidlink{0000-0002-2890-3725},}
\author[18,19,20]{{E.~Gaztañaga},}
\author[2]{{S.~Gontcho A Gontcho}\orcidlink{0000-0003-3142-233X},}
\author[21]{{C.~Howlett}\orcidlink{0000-0002-1081-9410},}
\author[22]{{D.~Kirkby}\orcidlink{0000-0002-8828-5463},}
\author[2]{{T.~Kisner}\orcidlink{0000-0003-3510-7134},}
\author[2]{{A.~Kremin}\orcidlink{0000-0001-6356-7424},}
\author[2]{{A.~Lambert},}
\author[2]{{M.~Landriau}\orcidlink{0000-0003-1838-8528},}
\author[23]{{J.~Lasker}\orcidlink{0000-0003-2999-4873},}
\author[24]{{L.~Le~Guillou}\orcidlink{0000-0001-7178-8868},}
\author[2]{{M.~E.~Levi}\orcidlink{0000-0003-1887-1018},}
\author[25,26]{{M.~Manera}\orcidlink{0000-0003-4962-8934},}
\author[6,7,8]{{P.~Martini}\orcidlink{0000-0002-4279-4182},}
\author[27]{{A.~Meisner}\orcidlink{0000-0002-1125-7384},}
\author[28,26]{{R.~Miquel},}
\author[29]{{J.~Moustakas}\orcidlink{0000-0002-2733-4559},}
\author[30]{{A.~D.~Myers},}
\author[31,32]{{G.~Niz}\orcidlink{0000-0002-1544-8946},}
\author[1,2]{{N.~Palanque-Delabrouille}\orcidlink{0000-0003-3188-784X},}
\author[33,34,35]{{W.~J.~Percival}\orcidlink{0000-0002-0644-5727},}
\author[2,36,4]{{C.~Poppett},}
\author[37]{{G.~Rossi},}
\author[38]{{E.~Sanchez}\orcidlink{0000-0002-9646-8198},}
\author[2]{{D.~Schlegel},}
\author[39,40]{{M.~Schubnell},}
\author[41]{{H.~Seo}\orcidlink{0000-0002-6588-3508},}
\author[27]{{D.~Sprayberry},}
\author[40]{{G.~Tarl\'{e}}\orcidlink{0000-0003-1704-0781},}
\author[12]{{M.~Vargas-Maga\~na}\orcidlink{0000-0003-3841-1836},}
\author[27]{{B.~A.~Weaver},}
\author[24]{{P.~Zarrouk}\orcidlink{0000-0002-7305-9578},}
\author[2]{{R.~Zhou}\orcidlink{0000-0001-5381-4372},}
\author[42]{{H.~Zou}\orcidlink{0000-0002-6684-3997}}

\abstract{
We present a method to mitigate the effects of fiber assignment incompleteness in two-point power spectrum and correlation function measurements from galaxy spectroscopic surveys, by truncating small angular scales from estimators. We derive the corresponding modified correlation function and power spectrum windows to account for the small angular scale truncation in the theory prediction. We validate this approach on simulations reproducing the Dark Energy Spectroscopic Instrument (DESI) Data Release 1 (DR1) with and without fiber assignment. We show that we recover unbiased cosmological constraints using small angular scale truncated estimators from simulations with fiber assignment incompleteness, with respect to standard estimators from complete simulations. Additionally, we present an approach to remove the sensitivity of the fits to high $k$ modes in the theoretical power spectrum, by applying a transformation to the data vector and window matrix. We find that our method efficiently mitigates the effect of fiber assignment incompleteness in two-point correlation function and power spectrum measurements, at low computational cost and with little statistical loss.
}

\begin{document}
\maketitle
\flushbottom

\section{Introduction}
\label{sec:intro}

Contemporary galaxy spectroscopic surveys, such as the Dark Energy Spectroscopic Instrument (DESI)~\cite{levi_desi_2013}, use thousands of optical fibers to collect spectra from millions of galaxies. Due to the increasing size of the collected data, cosmological measurements are becoming ever more precise and the impact of systematic effects is of growing importance. In particular, due to instrumental effects, spectroscopic surveys collect redshifts for only a fraction of the full set of targets available. For a given telescope pointing, two fibers cannot be positioned infinitely close to each other. In previous surveys such as the Sloan Digital Sky Survey (SDSS), this limitation was imposed by the size of the fiber mechanical holder. In the case of DESI, the patrol area of each fiber positioner is fixed by design. This limitation imposes a lower threshold on the angular separation between two observed galaxies. This induces a fiber assignment incompleteness effect that is uneven across the survey footprint and mostly affects regions with high density of targets. As a result, if not accounted for, missing galaxy pairs introduce a bias into the estimations of standard two-point statistics, namely the two-point correlation and two-point power spectrum, especially on small scales~\cite{hahn_effect_2017}. The overall effect of fiber assignment is reduced by having multiple visits at each position on the sky. In the final data release of the Baryon Oscillation Spectroscopic Survey (BOSS)~\cite{dawson_baryon_2013}, 5\% of targets were left unobserved as a result of fiber collisions~\cite{reid_sdss-iii_2016}. In DESI, the fiber incompleteness depends on the density of each target type, its priority in the fiber assignment process and the number of passes at each point of the survey. For the initial year of DESI observations (DESI DR1)~\cite{DESI2024.I.DR1}, the completeness ranges between 10\% in the Southern Galactic Cap to almost 100\% in the equatorial region of the Northern Galactic Cap.

Various methods have been developed to correct the effect of missing observations in galaxy clustering measurements~\cite{anderson_clustering_2012, guo_new_2012, hahn_effect_2017, pinol_imprint_2017, burden_mitigating_2017, bianchi_unbiased_2018, bianchi_confronting_2020}. The nearest neighbour method~\cite{anderson_clustering_2012} assigns the number of galaxies with missing redshift as a weight to the nearest observed galaxy in angular separation. This allows us to reconstruct the underlying two-point statistics on scales much larger than the fiber collision scale, but fails at smaller scales. Pairwise-inverse-probability (PIP) weighting, introduced in~\cite{bianchi_unbiased_2017, bianchi_unbiased_2018, bianchi_confronting_2020}, weights each galaxy pair by its inverse probability of being observed. It requires running the fiber assignment algorithm hundreds of times. This is perfectly feasible for the DESI samples but it can be challenging for processing thousands of mocks for systematic studies. Furthermore, this method provides unbiased clustering measurements only in regions of the survey footprint where pairs of galaxies have a non-zero probability of being observed and techniques based on angular up-weighting were developed to circumvent this issue~\cite{percival_using_2017, mohammad_completed_2020}. Other techniques propose to null purely angular modes~\cite{burden_mitigating_2017, paviot_angular_2022} that are affected either by fiber assignment incompleteness or by photometric systematics across the survey footprint. However, these techniques null all angular modes while, in the case of fiber assignment, we expect simple completeness weighting to recover large scale angular modes leaving only small angular scales to be corrected. \cite{zhao_completed_2021} proposed to subtract transverse modes from the power spectrum estimator, which we show does not sufficiently reduce the effect of fiber assignment incompleteness in appendix~\ref{sec:chained_multipoles}.

Another approach consists in approximating the effect of fiber collisions in two-point measurements by a top-hat function as a function of transverse separation, and convolving it with the true power spectrum in Fourier space to get a power spectrum model that includes the effect of missing observations~\cite{hahn_effect_2017}. This technique was successfully applied to eBOSS ELG~\cite{de_mattia_completed_2021} and quasar~\cite{neveux_completed_2020} samples. This is however an average correction that does not account for the local impact of the density of targets on the fraction of missing pairs. Working in configuration space, \cite{reid_25_2014} proposed to remove all pairs for which the angle between the line connecting the two galaxies of a pair and the line-of-sight is small. This was further investigated in~\cite{zarrouk_clustering_2018} for the clustering analysis of eBOSS quasars, where it improves upon nearest neighbor up-weighting in reducing the systematic bias. Such approach has the advantage of not relying on observed galaxies to infer properties of unobserved ones, at a very low statistical cost.

In this paper, we adopt a modified version of the previous method by removing all pairs of galaxies with angular separation below the fiber collision angular scale in configuration space for the correlation function estimator and in Fourier space for the power spectrum estimator (section~\ref{sec:desi_theta_cut}). We modify windows (relating the aforementioned estimators to the theory prediction) correspondingly in configuration and Fourier space to account for this truncation (section~\ref{sec:windows}). Finally, we validate our procedure with full-shape fits (section~\ref{sec:full_shape_fits}) and we discuss and reduce the sensitivity of the modified Fourier-space window to high-$k$ modes (section~\ref{sec:removing_sensitivity_to_high_k_theory}). Our methodology is used in the full-shape analysis of DESI DR1 data~\cite{DESI2024.V.KP5, DESI2024.VII.KP7B}.

\section{Mitigating fiber assignment incompleteness in two-point measurements}\label{sec:desi_theta_cut}

\subsection{Fiber assignment incompleteness in DESI}\label{sec:desi_fiber_collisions}

DESI is a robotic, fiber-fed, highly multiplexed spectroscopic survey that operates on the Mayall 4-meter telescope at Kitt Peak National Observatory~\cite{desi_collaboration_overview_2022}. DESI is currently conducting a five-year survey of about a third of the sky, and will obtain spectra for approximately 40 million galaxies and quasars~\cite{desi_collaboration_desi_2016}. 
Measurements of the baryon acoustic oscillations from the first year of DESI data have been released in~\cite{desi_collaboration_desi_2024, desi_collaboration_desi_2024-1} and the corresponding cosmological results are described in~\cite{desi_collaboration_desi_2024-2}. 
The survey includes a Bright Galaxy Survey (BGS) at $0.1 < z < 0.4$, Luminous Red Galaxies (LRGs) in the redshift range $0.4 < z < 1.1$, Emission Line Galaxies (ELGs) in the redshift range $0.8 < z < 1.6$, and quasars (QSOs) in the redshift range $0.8 < z < 2.1$, as well as the Lyman-$\alpha$ forest from QSO spectra with redshifts from 1.8 to 4.2. The largest sample in the full survey will be that of ELGs, with an expected number of about 18 million collected redshifts. Targets for these different tracers are assigned fibers through the fiber assignment algorithm \texttt{fiberassign}~\cite{FBA.Raichoor.2024}.\footnote{\url{https://github.com/desihub/fiberassign}.}

\begin{figure}
\centering
\includegraphics[scale=0.5]{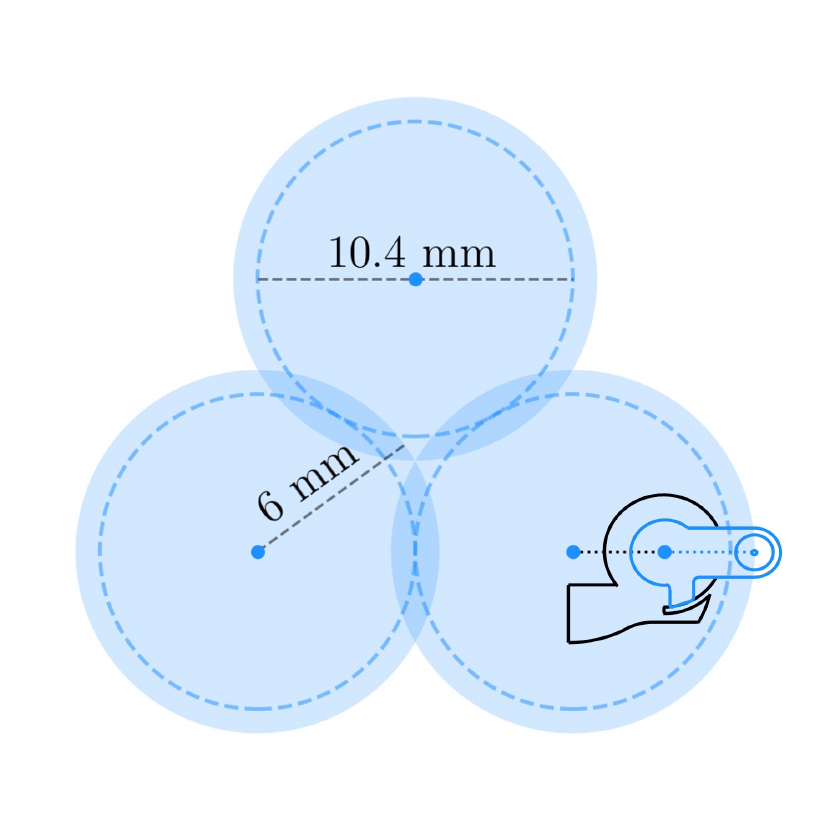}
\caption{Patrol areas of three neighbouring positioners (filled blue disks) in DESI focal plane~\cite{desi_collaboration_desi_2016-1}. The patrol diameter corresponds to a $\theta \sim 0.05^{\circ}$ angle on the sky (e.g. $\sim 2 \; \Mpc/h$ at redshift 1). A fully extended positioner is represented in solid lines, with the black and blue parts corresponding to the two rotating bodies of the positioner. The fiber has a 107 $\mu$m diameter core and is bonded into a 1.25 mm  ferrule (enclosed by the solid blue circle at the tip of the positioner)~\cite{silber_robotic_2023}.}
\label{fig:fiber_positioners_patrol}
\end{figure}

The DESI focal plane~\cite{silber_robotic_2023} incorporates about 5000 fiber positioners shown in figure 4.2 in~\cite{desi_collaboration_desi_2016-1}. Each fiber positioner is equipped with a robotic arm that can move the fiber within a fixed diameter of 12 mm, with neighbouring positioners' patrol area slightly overlapping, as seen in figure~\ref{fig:fiber_positioners_patrol}. Translating into angular separation, positioners cover a maximum angle varying from $0.0455^{\circ}$ to $0.0495^{\circ}$, depending on the distance to the center of the focal plane \cite{miller_optical_2023, anand-private-comm}. The DESI focal plane runs across the $\sim 14,000 \; \deg^2$ survey footprint through overlapping tiles --- one tile corresponding to one position of the instrument on the sky --- ensuring that at the end of the survey, each coordinate of the footprint will have up to seven visits (5.2 on average) for the dark time program and four visits (3.2 on average) for the bright time program~\cite{schlafly_survey_2023}. This configuration aims at recovering missing galaxies that would be overlooked with only a single layer of observations. In this work, we focus on the dataset collected during the initial year of DESI observations (DESI DR1), where only a fraction of targets have been observed ($\sim 35$\% of ELGs, $\sim 69$\% of LRGs, $\sim 87$\% of QSOs and $\sim 64$\% of BGS)~\cite{DESI2024.II.KP3}. This low completeness, and thus large fraction of missing small angular pairs ($\sim 50\%$, as seen below) led us to develop the method presented in this paper to account for the impact of fiber assignment incompleteness.

\subsection{N-body mock DESI catalogs}

We use 25 AbacusSummit N-body simulations~\cite{maksimova_abacussummit_2021, garrison_abacus_2021} to test our method. Each mock catalog is obtained from three simulations snapshots with $2 \; \Gpc/h$ side length reproducing the clustering of DESI dark time tracers (LRGs, ELGs and QSOs)
at three different redshifts. Halo occupation distribution (HOD) models were fitted on the data from the DESI Early Data Release~\cite{desi_collaboration_validation_2024, desi_collaboration_early_2023}. HOD models for LRGs and QSOs are described in~\cite{yuan_desi_2024}, and HODs for ELGs in~\cite{rocher_desi_2023}. Cubic mocks are replicated and cut to reproduce the DESI footprint. The radial selection function, which is measured through the DESI spectroscopic pipeline~\cite{guy_spectroscopic_2023}, and imaging masks of DR1 data are then applied. The resulting mocks are called cut-sky mocks. In order to study the effect of fiber assignment on two-point measurements, we use so-called \emph{complete} cut-sky mocks, that include all targets reachable by a fiber, as well as \emph{\altmtl} (alternate Merged Target Ledger) \cite{KP3s7-Lasker} mocks. In the latter, the tiling sequence and hardware status of each fiber channel from the actual DESI DR1 observations~\cite{schlafly_survey_2023} are used as input of the fiber assignment \texttt{fiberassign} algorithm.

\subsection{Effect of fiber assignment incompleteness on the clustering of DESI DR1 samples}\label{sec:fiber_assign_incomp_effect}

\begin{figure}[ht]
\begin{center}
\includegraphics[scale=0.8]{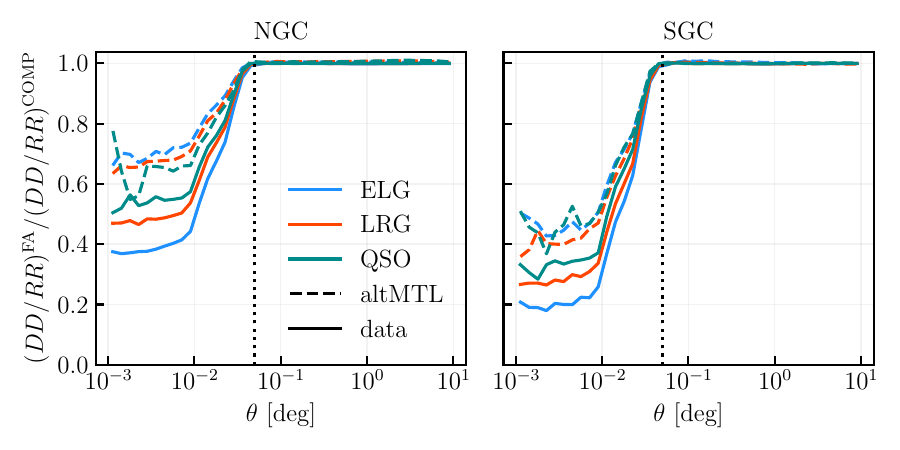}
\end{center}
\caption{Fraction of observed galaxy pair counts in DESI DR1 fiber-assigned data (plain line) and in one \altmtl mock (dashed line) with respect to the corresponding complete catalog of targets, as a function of $\theta$. Incompleteness at large angular separation in data and \altmtl\ mocks has been corrected with completeness weights~\cite{DESI2024.II.KP3}. The 3 DESI dark time tracers - ELGs, LRGs, QSOs - and the two galactic caps are shown. $(DD/RR)^{\mathrm{FA}}$ (resp. $(DD/RR)^{\mathrm{COMP}}$) refers to pair counts of fiber-assigned (resp. complete) data or mock, normalized by pair counts of random points sampled within the same footprint and selection function as the data. Vertical dotted lines indicate $\Lambda_{\theta} = 0.05^{\circ}$.}
\label{fig:dd_fiber_collisions}
\end{figure}

Fiber assignment incompleteness has two main effects on observed clustering in DESI. First, on angular scales larger than the patrol diameter, the density of targets that can be observed with one tile is capped in a way that depends on the details of fiber assignment (requirements on sky fibers, standard stars, subpriorities, etc.). To correct for that, we apply completeness weights, calculated at the level of each fiber and target class, that are presented in \cite{DESI2024.II.KP3} and \cite{KP3s15-Ross}. Second, below the patrol diameter angular scale, there are missing pairs that cannot be recovered with completeness weights. The aim of the method we are presenting in this paper is to mitigate this effect. Figure~\ref{fig:dd_fiber_collisions} shows the fraction of galaxy pair counts in \altmtl\ mocks with respect to complete mocks, together with that of fiber-assigned targets with respect to all targets, as a function of angular separation $\theta$. The survey is divided into the northern galactic cap (NGC) and southern galactic cap (SGC). As seen in figure~\ref{fig:dd_fiber_collisions}, SGC is less complete than NGC as a consequence of the survey strategy and observational conditions during the first year of DESI operations. Moreover, the ELG sample has the lowest completeness, with its ratio of fiber-assigned to complete pair counts dropping to about 20\% at small $\theta$ in SGC. In the following, we restrict ourselves to the ELG sample to test our method to mitigate the effect of fiber assignment incompleteness. In addition, we consider the $1.1 < z < 1.6$ redshift bin, as used in the DESI DR1 fiducial analysis~\cite{DESI2024.II.KP3}.

From figure~\ref{fig:dd_fiber_collisions} we see that fiber assignment incompleteness affects clustering below $\theta \sim 0.05^{\circ}$ for all samples. This is consistent with the DESI fiber positioner patrol diameter mentioned in section~\ref{sec:desi_fiber_collisions}. Furthermore, because of their physical dimensions (as can be seen on figure~\ref{fig:fiber_positioners_patrol}) there is an additional effect arising from "collisions" between neighboring positioners that would affect smaller angular separations $\theta \sim 0.01^{\circ}$. 
Thus, for the remainder of this paper, we choose a threshold $\Lambda_{\theta} = 0.05^{\circ}$ to discard pair counts at $\theta < \Lambda_{\theta}$ in order to remove the effect of fiber assignment incompleteness.
We note some difference between \altmtl\ mocks and fiber-assigned data, further discussed in~\cite{KP3s6-Bianchi}. The proposed $\Lambda_{\theta}$ cut makes the analysis robust to the forward modelling of the fiber assignment in mock catalogs. 

\subsection{\texorpdfstring{$\theta$}{theta}-cut correlation function}\label{sec:corr_estimator}

\begin{figure}
\centering
\includegraphics[scale=0.78]{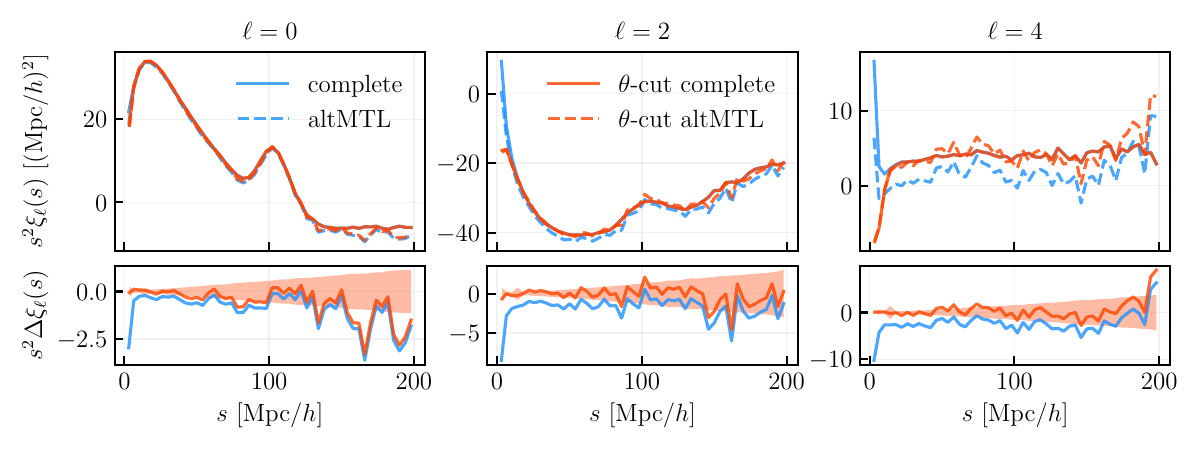}
\caption{{\it Top:}
average correlation function multipoles ($\ell$ = 0, 2, 4) of 25 DR1 ELG complete (solid lines) and \altmtl\ (dashed lines) AbacusSummit mocks, in the redshift bin $1.1 < z < 1.6$. Blue (resp. red) curves stand for the correlation function without (resp. with) the $\theta$-cut (i.e. removing pair counts at $\theta < 0.05^{\circ}$).
{\it Bottom:} difference between multipoles of \altmtl\ and complete mocks, with and without 
cutting out pair counts at $\theta < 0.05^{\circ}$.
The shaded area indicates 1/5th of the standard deviation of DESI DR1 correlation function, estimated from approximate mocks (see section~\ref{sec:EZmocks_covariance}).}
\label{fig:corr_fiberassign_thetacut0.05_ELG_25mocks}
\end{figure}

In this section, we write an estimator for the correlation function with pair counts at $\theta < \Lambda_{\theta}$ removed, which we call $\theta$-cut correlation function. In the following, $D$ stands for galaxies from the data sample and $R$ for galaxies from a random sample with the same selection function as the data. The standard Landy-Szalay estimator~\cite{landy_bias_1993} for the correlation function multipole of order $\ell$ reads:
\begin{equation}
\widehat{\xi}_{\ell}(s) = \frac{2 \ell + 1}{|\lbrace\mu, RR \neq 0 \rbrace|} \times \int_{\mu, RR \neq 0} \frac{DD(s, \mu) - DR(s, \mu) - RD(s, \mu) + RR(s, \mu)}{RR(s, \mu)} \Leg{\ell}(\mu) d\mu.
\label{eq:correlation_function_estimator}
\end{equation}
where $s$ is the comoving pair separation, $\mu$ is the cosine of the angle between the vector formed by the pair of galaxies $\vs$ and the mid-point line-of-sight $(\vr_{1} + \vr_{2})/2$, and $DD(s, \mu)$, $DR(s, \mu)$, and $RR(s, \mu)$ are data - data, data - random, random - random weighted pair counts in a given $s$ and $\mu$-bin, respectively. Pairs are weighted by the product of the individual weights of the two
objects in the pair, which themselves are the product of systematic correction weights (see~\cite{DESI2024.II.KP3}) and FKP weights~\cite{feldman_power-spectrum_1994}.
In practice, $\int_{\mu, RR \neq 0}$ denotes a (finite) sum over $\mu$-bins for which $RR(s, \mu) \neq 0$, $|\lbrace\mu, RR \neq 0 \rbrace|$ is the sum of the width of corresponding $\mu$-bins, and $\Leg{\ell}(\mu) d\mu$ is taken as the integral of Legendre multipole of order $\ell$ over each $\mu$-bin.\footnote{This convention ensures that $\widehat{\xi}_{\ell}(s) = 0$ for $\ell > 0$ when the integrand does not depend on $\mu$.} Let us write an estimator of the $\theta$-cut correlation function multipole of order $\ell$, $\hat{\xi}_{\ell}^{\cut}$:
\begin{equation}
\begin{split}
    \widehat{\xi}_{\ell}^{\cut}(s) 
    &= \frac{2 \ell + 1}{|\lbrace\mu, RR^{\cut} \neq 0 \rbrace|} \\ &\times \int_{RR^{\cut} \neq 0} d\mu \frac{DD^{\cut}(s, \mu) - DR^{\cut}(s, \mu)-RD^{\cut}(s, \mu) + RR^{\cut}(s, \mu)}{RR^{\cut}(s, \mu)} \Leg{\ell}(\mu).
\end{split}
\label{eq:cut_correlation_function_estimator}
\end{equation}
where $DD^{\cut}(s, \mu)$ are the data - data weighted counts, removing pairs of angular separations $\theta < \Lambda_{\theta}$ (similarly for $DR^{\cut}$ and $RR^{\cut}$). In practice we use the Python package \texttt{pycorr}\footnote{\url{https://github.com/cosmodesi/pycorr}, wrapping a modified version of \texttt{Corrfunc} \url{https://github.com/manodeep/Corrfunc}~\cite{sinha_corrfunc_2020}.} to compute the correlation function. We use 200 $\mu$-bins between $-1$ and $1$, and $s$-bins of width $4 \, \mathrm{Mpc}/h$. Figure~\ref{fig:corr_fiberassign_thetacut0.05_ELG_25mocks} shows the average correlation function multipoles of the 25 complete and \altmtl\ AbacusSummit mocks, with and without cutting out $\theta < \Lambda_{\theta}$ galaxy pairs. The $\theta$-cut correlation function is estimated using equation~\eqref{eq:cut_correlation_function_estimator}. The shaded area indicates 1/5th of the standard deviation of DESI DR1 correlation function multipoles, which is estimated from 1000 approximate mocks cut to the DESI DR1 footprint and processed through fast fiber assignment (see section~\ref{sec:EZmocks_covariance}). Figure~\ref{fig:corr_fiberassign_thetacut0.05_ELG_25mocks} shows that using estimator~\eqref{eq:cut_correlation_function_estimator}, the average $\theta$-cut correlation function of the 25 fiber-assigned mocks is in agreement with that of complete mocks within about 1/5th of DESI DR1 precision. Modifying the standard estimator with the $\theta$-cut allows us to remove the effect of missing galaxy close pairs from the correlation function, at the expense of changing the theory as we explain in section~\ref{sec:window_corr}.

\subsection{\texorpdfstring{$\theta$}{theta}-cut power spectrum}\label{sec:power_estimator}

Let us write an estimator for the $\theta$-cut power spectrum. The standard two-point power spectrum estimator~\cite{yamamoto_measurement_2006} is based on the FKP field:
\begin{equation}
F(\vr) = n_{D}(\vr) - \alpha_R n_{R}(\vr).
\label{eq:fkp}
\end{equation}
where $n_{D}(\vr)$ and $n_{R}(\vr)$ are 
the densities at comoving position $\vr$ of the observed galaxies and random catalog, respectively.
Both densities include weights to correct for various systematic effects (see~\cite{DESI2024.II.KP3}) and to optimize variance (FKP weights~\cite{feldman_power-spectrum_1994}): $w_D$, $w_R$. $\alpha_R$ is the ratio of the overall weight of galaxies to that of randoms. The power spectrum multipoles are given by~\cite{bianchi_measuring_2015}:
\begin{equation}
P_{\ell}(k) = \frac{2 \ell + 1}{A} \int \frac{d\Omega_{k}}{4 \pi} F_{0}(\vk) F_{\ell}(-\vk) - \mathcal{N}_{\ell}
\label{eq:power_spectrum_multipoles}
\end{equation}

with:
\begin{equation}
F_{\ell}(\vk) = \int d\vr F(\vr) \mathcal{L}_{\ell}(\hat{\vk} \cdot \hat{\eta}) e^{i \vk \cdot \vr}
\label{eq:fkp_multipoles}
\end{equation}
where hat quantities denote vectors normalized to unity. The line-of-sight $\hat{\eta}$ is taken to be $\hat{\vr}_1$, the position of the first galaxy of the pair with respect to the observer. The normalization term is estimated as $A = \alpha_R \int d\vr n_{R}(\vr) n_{D}(\vr)$. $\int \frac{d\Omega_{k}}{4 \pi}$ denotes the spherical average in a $k$-bin, and $\int d\vr$ are to be understood as a sum over a 3D-mesh onto which $n_R(\vr)$, $n_D(\vr)$ and thus $F(\vr)$ are estimated. The shot noise term $\mathcal{N}_{\ell}$ is only non-zero for the monopole:
\begin{equation}
\mathcal{N}_{0} = \frac{1}{A} \left[ \sum_{i=1}^{N_D} w_{D,i}^2 + \alpha_{R}^2 \sum_{i=1}^{N_R} w_{R,i}^2 \right]
\label{eq:pk_shotnoise}
\end{equation}
where the $\sum_{i=1}^{N_D}$ (resp. $\sum_{i=1}^{N_R}$) runs over objects in the data (resp. randoms) catalog. 
The $\theta$-cut power spectrum estimator is obtained by subtracting, from this estimator, the contribution from low-$\theta$ pairs: 
\begin{equation}
    \widehat{P}_{\ell}^{\mathrm{cut}}(k) = \widehat{P}_{\ell}(k) - \widehat{\Delta P}_{\ell}(k).
\label{eq:cut_power_spectrum_estimator}
\end{equation}
To estimate $\widehat{\Delta P}_{\ell}(k)$, we use a direct pair counting approach (similarly to equation 26 in~\cite{bianchi_confronting_2020}), appropriately scaling randoms to data density, and removing contributions from self-pairs in case of autocorrelation:
\begin{equation}
    \widehat{\Delta P}_{\ell}(k) = \widehat{\Delta P}_{\ell}^{D D}(k) - \widehat{\Delta P}_{\ell}^{D R}(k) - \widehat{\Delta P}_{\ell}^{R D}(k) + \widehat{\Delta P}_{\ell}^{R R}(k)
\label{eq:power_pair_counting_estimate}
\end{equation}
 with:
\begin{equation}
\widehat{\Delta P}_{\ell}^{X Y}(k) = i^{\ell} \frac{2 \ell + 1}{A} \sum_{i, j, \; \theta \leq \Lambda_{\theta}} w_{X, i} w_{Y, j} j_{\ell}(k |\vr_{j} - \vr_{i}|) \Leg{\ell}(\hat{\vr}_{i} \cdot \widehat{\vr_{j} - \vr_{i}})
\label{eq:power_direct}
\end{equation}
where $w_{X, i}$, $w_{Y, j}$ are particle weights, the sum $\sum_{i, j, \; \theta \leq \Lambda_{\theta}}$ runs over pairs of particles with angular separation $\theta \leq \Lambda_{\theta}$, and $j_{\ell}$ are spherical Bessel functions of the first kind. We compute the power spectrum with \texttt{pypower}.\footnote{\url{https://github.com/cosmodesi/pypower}.} We use a random catalog of size about 27 times larger than that of each mock catalog. The FKP field~\eqref{eq:fkp} is computed in a box of side 9 Gpc/$h$ encapsulating the ELG catalog volume, and interpolated on a mesh of 6 Mpc/$h$ cells, with the triangular shape cloud (TSC) scheme. Its Fourier-transformed multipoles~\eqref{eq:fkp_multipoles} are computed using a Fast Fourier Transform (FFT) algorithm and interlacing of order 3 to reduce aliasing effects~\cite{sefusatti_accurate_2016}. We use $k$-bins of width $\Delta k = 0.005 \; h/\mathrm{Mpc}$.
In appendix~\ref{sec:thetacut_powerspectrum_estimator_numerical_details}, we discuss the numerical implementation of equation~\eqref{eq:power_direct}, and show that it can be accurately computed within a few minutes for one mock on one Perlmutter CPU node on NERSC.

\begin{figure}
\centering
\includegraphics[scale=0.78]{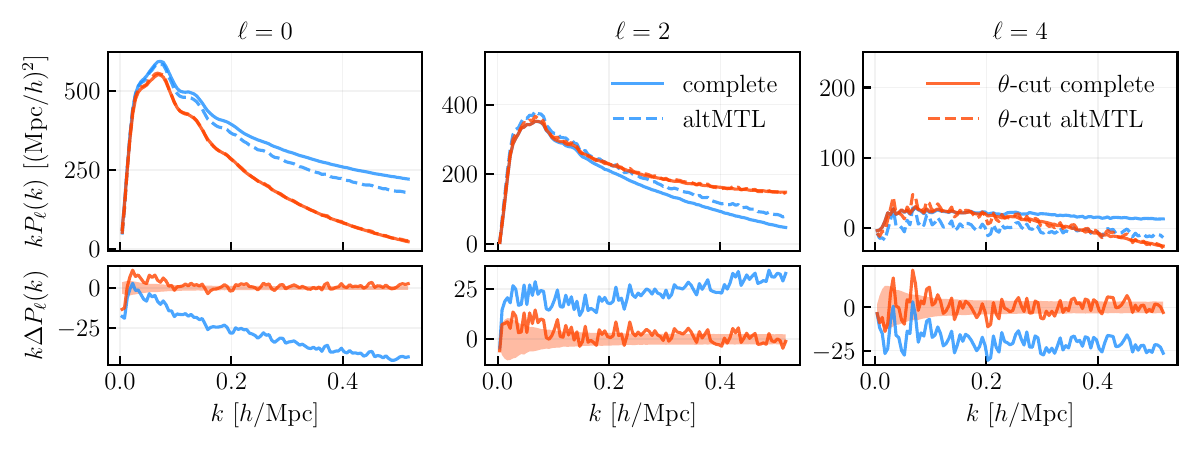}
\caption{{\it Top:}
average power spectrum multipoles ($\ell$ = 0, 2, 4) of the 25 DR1 ELG complete (solid lines) and \altmtl\ (dashed lines) AbacusSummit mocks in the redshift bin $1.1 < z < 1.6$. Red (resp. blue) curves stand for the power spectrum with (resp. without) the $\theta$-cut 
(i.e. removing pair counts at $\theta < 0.05^{\circ}$).
{\it Bottom:}
difference between multipoles of \altmtl\ and complete mocks, with and without
cutting out pair counts at $\theta < 0.05^{\circ}$.
The shaded area indicates 1/5th of the standard deviation of DESI
DR1 power spectrum, estimated from approximate mocks (see section~\ref{sec:EZmocks_covariance}).
}
\label{fig:power_fiberassign_thetacut0.05_ELG_25mocks}
\end{figure}

Figure~\ref{fig:power_fiberassign_thetacut0.05_ELG_25mocks} shows the average power spectrum multipoles of the 25 AbacusSummit mocks (in the redshift bin $1.1 < z < 1.6$), in the complete and \altmtl\ cases, with and without $\theta$-cut. $\theta$-cut power spectrum multipoles are computed using~\eqref{eq:cut_power_spectrum_estimator}.
Shaded areas on the bottom panels show 1/5th of DESI DR1 standard deviation, estimated from approximate mocks (see section~\ref{sec:EZmocks_covariance}).
Figure~\ref{fig:power_fiberassign_thetacut0.05_ELG_25mocks} shows that using the above estimator, the $\theta$-cut power spectrum of fiber-assigned mocks is in agreement with that of complete mocks within about 1/5th of DESI DR1 precision. 
As for the correlation function, while the $\theta$-cut allows us to suppress the effect of missing galaxy close pairs, we need to account for this modification of the standard power spectrum estimator in the model.

\section{Window matrix for \texorpdfstring{$\theta$}{theta}-cut estimator modelling}\label{sec:windows}

When comparing the correlation function or power spectrum measurement from a survey to their theoretical prediction, one needs to properly model the survey footprint and selection function in order to get unbiased cosmological constraints. This is usually done through a so-called window matrix that includes the geometrical features of the survey and relates the expected value of the measurement to the theory. With $\theta$-cut statistics, we are adding a geometrical effect that needs to be modelled in the window matrix. This section presents how we modify standard windows to incorporate this cut.

\subsection{Correlation function window}\label{sec:window_corr}
In general, the window matrix links the theory correlation function multipoles to the expected value of observed correlation multipoles through:
\begin{equation}
    \aver{\widehat{\xi}_{\ell}(s)} = W_{\ell \ell^{\prime}}(s) \xi_{\ell^{\prime}}(s).
\end{equation}
Let us write the $\theta$-cut window matrix $W_{\ell \ell^{\prime}}^{\cut}(s)$. The expected value for our implementation of the Landy-Szalay estimator in equation~\eqref{eq:correlation_function_estimator} is:
\begin{equation}
\begin{split}
\aver{\widehat{\xi}_{\ell}^{\cut}(s)}
&= \frac{2 \ell + 1}{|\lbrace\mu, RR^{\cut} \neq 0 \rbrace|} \\
&\times \int_{RR^{\cut} \neq 0} d\mu \aver{\frac{DD^{\cut}(s, \mu) - DR^{\cut}(s, \mu) - RD^{\cut}(s, \mu) + RR^{\cut}(s, \mu)}{RR^{\cut}(s, \mu)}} \Leg{\ell}(\mu) \\
&\simeq \frac{2 \ell + 1}{|\lbrace\mu, RR^{\cut} \neq 0 \rbrace|} \int_{RR^{\cut} \neq 0} d\mu \frac{RR^{\cut}(s, \mu) \xi(s, \mu)}{RR^{\cut}(s, \mu)} \Leg{\ell}(\mu) \\
&= \frac{2 \ell + 1}{|\lbrace\mu, RR^{\cut} \neq 0 \rbrace|} \int_{RR^{\cut} \neq 0} d\mu \xi(s, \mu) \Leg{\ell}(\mu)
\end{split}
\end{equation}
where $\xi(s, \mu)$ is the theory correlation function.

Then, expanding $\xi(s, \mu)$ into Legendre multipoles, we get:
\begin{equation}
\aver{\widehat{\xi}_{\ell}^{\cut}(s)} = \frac{2 \ell + 1}{|\lbrace\mu, RR^{\cut} \neq 0 \rbrace|} \int_{RR^{\cut} \neq 0} d\mu \sum_{\ell^{\prime}} \xi_{\ell^\prime}(s) \Leg{\ell^{\prime}}(\mu) \Leg{\ell}(\mu)
\label{eq:corr_theory}
\end{equation}
i.e.:
\begin{equation}
\aver{\widehat{\xi}_{\ell}^{\cut}(s)} = \sum_{\ell^{\prime}} W_{\ell \ell^{\prime}}^{\cut}(s) \xi_{\ell^{\prime}}(s)
\end{equation}
with:
\begin{equation}
W_{\ell \ell^{\prime}}^{\cut}(s) = \frac{2 \ell + 1}{|\lbrace\mu, RR^{\cut} \neq 0 \rbrace|} \int_{RR^{\cut} \neq 0} d\mu \Leg{\ell^{\prime}}(\mu) \Leg{\ell}(\mu).
\label{eq:window_cut_correlation}
\end{equation}
In addition to the pure $\mu$-window above, we also account for the binning of the correlation function, effectively rebinning the theory $\xi_{\ell}(s)$ computed with $s$-theory bins of $1 \; \mathrm{Mpc}/h$ to the measurement binning of $4 \; \mathrm{Mpc}/h$, with weights $RR^{\cut}$.

\subsection{Power spectrum window}\label{sec:power_window}
Now let us write the $\theta$-cut power spectrum theory. Analogous to the correlation function, the observed power spectrum is linked to the theory through its window matrix. Injecting equation~\eqref{eq:fkp_multipoles} into~\eqref{eq:power_spectrum_multipoles}, we get:
\begin{equation}
\widehat{P}_{\ell}(k) = \frac{2 \ell + 1}{A} \int \frac{d\Omega_{k}}{4 \pi} \int d\vr \int d\vs F(\vr) F(\vr + \vs) e^{i \vk \cdot \vs} \Leg{\ell}(\hat{\vr} \cdot \hat{\vk}) - \mathcal{N}_{\ell}.
\label{eq:standard_power_estimator}
\end{equation}
and using that~\cite{feldman_power-spectrum_1994} (with $w(\vr)$ particle weights):
\begin{equation}
\aver{F(\vr) F(\vr + \vs)} = \bar{n}(\vr) \bar{n}(\vr + \vs) \xi(\vs) + w(\vr)\bar{n}(\vr)\delta_{D}^{(3)}(\vs),
\end{equation}
the power spectrum expectation value reads ($\mathcal{N}_{\ell}$ given by equation~\eqref{eq:pk_shotnoise} cancelling the above shot noise contribution $w(\vr)\bar{n}(\vr)\delta_{D}^{(3)}(\vs)$):
\begin{equation}
\aver{\widehat{P}_{\ell}(k)} = \frac{2 \ell + 1}{A} \int \frac{d\Omega_k}{4\pi} \int d\vr \int d\vs \bar{n}(\vr) \bar{n}(\vr + \vs) \xi(\vs) e^{i \vk \cdot \vs} \Leg{\ell}(\hat{\vr} \cdot \hat{\vk}).
\label{eq:power}
\end{equation}
Using the Rayleigh plane wave expansion:
\begin{equation}
e^{i \vk \cdot \vs} = \sum_{q=0}^{+\infty} i^q (2q + 1) j_{q}(ks) \mathcal{L}_{q}(\hat{\vk} \cdot \hat{\vs}) 
\end{equation}
and (for sufficiently fine $\hat{\vk}$ sampling of the power spectrum estimator mesh):
\begin{equation}
\int \frac{d\Omega_k}{4\pi} \mathcal{L}_{\ell}(\hat{\vk} \cdot \hat{\vr}) \mathcal{L}_{q}(\hat{\vk} \cdot \hat{\vs}) = \frac{\delta_{\ell q}}{2 \ell + 1} \mathcal{L}_{\ell}(\hat{\vr} \cdot \hat{\vs}),
\end{equation}
we find:
\begin{equation}
\aver{\widehat{P}_{\ell}(k)} = i^{\ell} \frac{2 \ell + 1}{A} \int d\vr \int d\vs \bar{n}(\vr) \bar{n}(\vr + \vs) \xi(\vs) j_{\ell}(k s) \Leg{\ell}(\hat{\vr} \cdot \hat{\vs})
\end{equation}
i.e.
\begin{equation}
\aver{\widehat{P}_{\ell}(k)} = i^{\ell} \frac{2 \ell + 1}{A} \int d\vr \int s^{2} ds \int d\phi \int d\mu \bar{n}(\vr) \bar{n}(\vr + \vs) \xi(\vs) j_{\ell}(k s) \Leg{\ell}(\mu)
\end{equation}
with $\mu = \hat{\vr} \cdot \hat{\vs}$ and $\phi$ the azimuthal angle of $\hat{\vs}$ around $\hat{\vr}$. Injecting the multipole expansion for $\xi(\vs)$ we have:
\begin{equation}
\aver{\widehat{P}_{\ell}(k)} = i^{\ell} \frac{2 \ell + 1}{A} \int d\vr \int s^{2} ds \int d\phi \int d\mu \bar{n}(\vr) \bar{n}(\vr + \vs) \sum_{\ell^{\prime}} \xi_{\ell^{\prime}}(s) \mathcal{L}_{\ell^{\prime}}(\mu) j_{\ell}(k s) \Leg{\ell}(\mu).
\end{equation}
In the previous equations, we have made the approximation that theory multipoles $\xi_{\ell^{\prime}}(s)$ are uniquely defined by the line-of-sight chosen as $\vr$, the position of the first galaxy of the pair with respect to us, which is only valid in the local plane-parallel approximation, $s \ll r$. This assumption leads to so-called wide-angle effects, that appear in wide-angle surveys where the maximum separation between two galaxies can be of the same order of magnitude as their distance to the observer $r$. To correct for that, $\xi(\vs)$ can be expanded around in powers of $s/r$~\cite{castorina_beyond_2018}:
\begin{equation}
\xi(\vs) = \sum_{\ell, n} \left( \frac{s}{r} \right)^{n} \xi_{\ell}^{(n)}(s) \mathcal{L}_{\ell} (\hat{\vr} \cdot \hat{\vs})
\label{eq:corr_wa}
\end{equation}
where:
\begin{equation}
\xi_{\ell}^{(n)}(s) = \frac{(-i)^{\ell}}{2 \pi^{2}} \int dk k^{2} (k s)^{-n} P_{\ell}^{(n)}(k) j_{\ell}(ks)
\label{eq:xi_to_pk}
\end{equation}
where $P_{\ell}^{(n)}$ (with $n=1$) is defined in appendix~\ref{sec:wide_angle_odd} (see below). Eventually the expected value of the power spectrum estimator reads:
\begin{equation}
\aver{\widehat{P}_{\ell}(k)} = i^{\ell} \frac{2 \ell + 1}{A} \int d\vr \int s^{2} ds \int d\phi \int d\mu \bar{n}(\vr) \bar{n}(\vr + \vs) \sum_{n, \ell^{\prime}} \left( \frac{s}{r} \right)^{n} \xi_{\ell^{\prime}}^{(n)}(s) \Leg{\ell^{\prime}}(\mu) j_{\ell}(k s) \Leg{\ell}(\mu)
\end{equation}
i.e.:
\begin{equation}
\aver{\widehat{P}_{\ell}(k)} = 4 \pi \int {k^{\prime}}^{2 - n} dk^{\prime} \sum_{n, \ell^{\prime}} W_{\ell \ell^{\prime}}^{(n)}(k, k^{\prime}) P_{\ell^{\prime}}^{(n)}(k^{\prime})
\label{eq:power_theory}
\end{equation}
where:
\begin{equation}
W_{\ell \ell^{\prime}}^{(n)}(k, k^{\prime}) = \frac{2 \ell + 1}{(2 \pi)^{3} A} i^{\ell} (-i)^{\ell^{\prime}} \int d\vr \int s^{2} ds \int d\phi \int d\mu \bar{n}(\vr) \bar{n}(\vr + \vs) r^{-n} j_{\ell^{\prime}}(k^{\prime} s) \Leg{\ell^{\prime}}(\mu) j_{\ell}(k s) \Leg{\ell}(\mu).
\end{equation}
We can rewrite $W_{\ell \ell^{\prime}}^{(n)}(k, k^{\prime})$:
\begin{equation}
W_{\ell \ell^{\prime}}^{(n)}(k, k^{\prime}) = \frac{i^{\ell} (-i)^{\ell^{\prime}}}{2 \pi^{2}}  \int s^{2} ds \sum_{p} \frac{2 \ell + 1}{2 p + 1} A_{p \ell \ell^{\prime}} W_{p}^{(n)}(s) j_{\ell^{\prime}}(k^{\prime} s) j_{\ell}(k s)
\label{eq:window_matrix}
\end{equation}
with $A_{p \ell \ell^{\prime}}$ defined by:
\begin{equation}
\Leg{\ell}(\mu) \Leg{\ell^{\prime}}(\mu) = \sum_{p} A_{p \ell \ell^{\prime}} \Leg{p}(\mu)
\end{equation}
and where $W_{p}^{(n)}(s)$ is the configuration space window function, defined by:
\begin{equation}
W_{p}^{(n)}(s) = \frac{2 p + 1}{4 \pi A}\int d\vr \int d\phi \int d\mu r^{-n} \bar{n}(\vr) \bar{n}(\vr + \vs) \Leg{p}(\mu).
\label{eq:power_window_function}
\end{equation}
Equation~\eqref{eq:power_window_function} is equal to the Hankel transform of the power spectrum of the random catalog, that can be estimated through the standard power spectrum estimator from equation~\eqref{eq:standard_power_estimator} (see appendix E.1 in~\cite{beutler_interpreting_2019}). Equations~\eqref{eq:power_theory} to \eqref{eq:power_window_function} hold in the standard case, with no $\theta$-cut. 

To obtain the $\theta$-cut window function, we just remove the contribution $\Delta W_{p}^{(n)}(s)$ of pairs at small angular separation:
\begin{equation}
    W_{p}^{\cut, (n)}(s) = W_{p}^{(n)}(s) - \Delta W_{p}^{(n)}(s)
\end{equation}
where $\Delta W_{p}^{(n)}(s)$ can be estimated with a direct pair counting approach:

\begin{equation}
\Delta W_{p}^{(n)}(s) = \frac{2 \ell + 1}{A} \sum_{i, j, \; \theta \leq \Lambda_{\theta}} r_{i}^{-n} w_{R, i} w_{R, j} \Leg{\ell}(\hat{\vr}_{i} \cdot \widehat{\vr_{j} - \vr_{i}})
\label{eq:window_direct}
\end{equation}

Eventually, one can relate theory $P_{\ell^{\prime}}^{(n)}(k^{\prime})$ of order $n = 1$ that enters equation~\eqref{eq:power_theory} to the plane-parallel multipoles $P_{\ell^{\prime}}^{(0)}(k^{\prime})$ linearly through $P_{\ell}^{(1)}(k) = \sum_{\ell^{\prime}}W_{\ell \ell^\prime}^{\mathrm{WA}, (1)}(k) P_{\ell^\prime}^{(0)}(k)$
with (see appendix~\ref{sec:wide_angle_odd}):
\begin{equation}
W_{\ell \ell^\prime}^{\mathrm{WA}, (1)}(k) = - i \frac{\ell \left(\ell - 1\right)}{2 \left(2 \ell - 1\right)} \left[ \left(\ell - 1\right) - k \partial_{k} \right] \delta_{\ell - 1, \ell^\prime} - i \frac{\left(\ell + 1\right) \left(\ell + 2\right)}{2 \left(2 \ell + 3\right)} \left[ \left(\ell + 2\right) + k \partial_{k} \right] \delta_{\ell + 1, \ell^\prime}
\end{equation}
where $\partial_{k}$ is the derivative operator w.r.t. $k$. We can thus include wide-angle correction into the window:
\begin{equation}
W_{\ell \ell^{\prime}}(k, k^{\prime}) = W_{\ell \ell^{\prime}}^{(0)}(k, k^{\prime}) + \frac{1}{k^\prime} \sum_{\ell^{\prime\prime}} W_{\ell \ell^{\prime\prime}}^{(1)}(k, k^{\prime}) W_{\ell^{\prime\prime} \ell^{\prime}}^{\mathrm{WA}, (1)}(k^{\prime})
\label{eq:window_matrix_wa}
\end{equation}
such that $\aver{\widehat{P}_{\ell}(k)} = 4 \pi \int {k^{\prime}}^{2} dk^{\prime} \sum_{\ell^{\prime}} W_{\ell \ell^{\prime}}(k, k^{\prime}) P_{\ell^{\prime}}^{(0)}(k^{\prime})$.

In the following, plots of the \textit{window matrix} show the quantity $4 \pi k^{\prime 2} dk^{\prime} W_{\ell \ell^{\prime}}(k, k^{\prime})$ (the integral over $k^{\prime}$ is replaced by a discrete sum over the $k^{\prime}$ bins, in the range $[0.001, 0.35] \; h/\Mpc$). Hence the window matrix has nine blocks, corresponding to each combination of theory and data multipoles, and each block has as many columns as the number of $k^{\prime}$ bins, and as many rows as the number of $k$ bins. Practical implementation of the window computation is discussed in section~\ref{sec:window}.

\begin{figure}
\begin{center}
\includegraphics[scale=0.8]{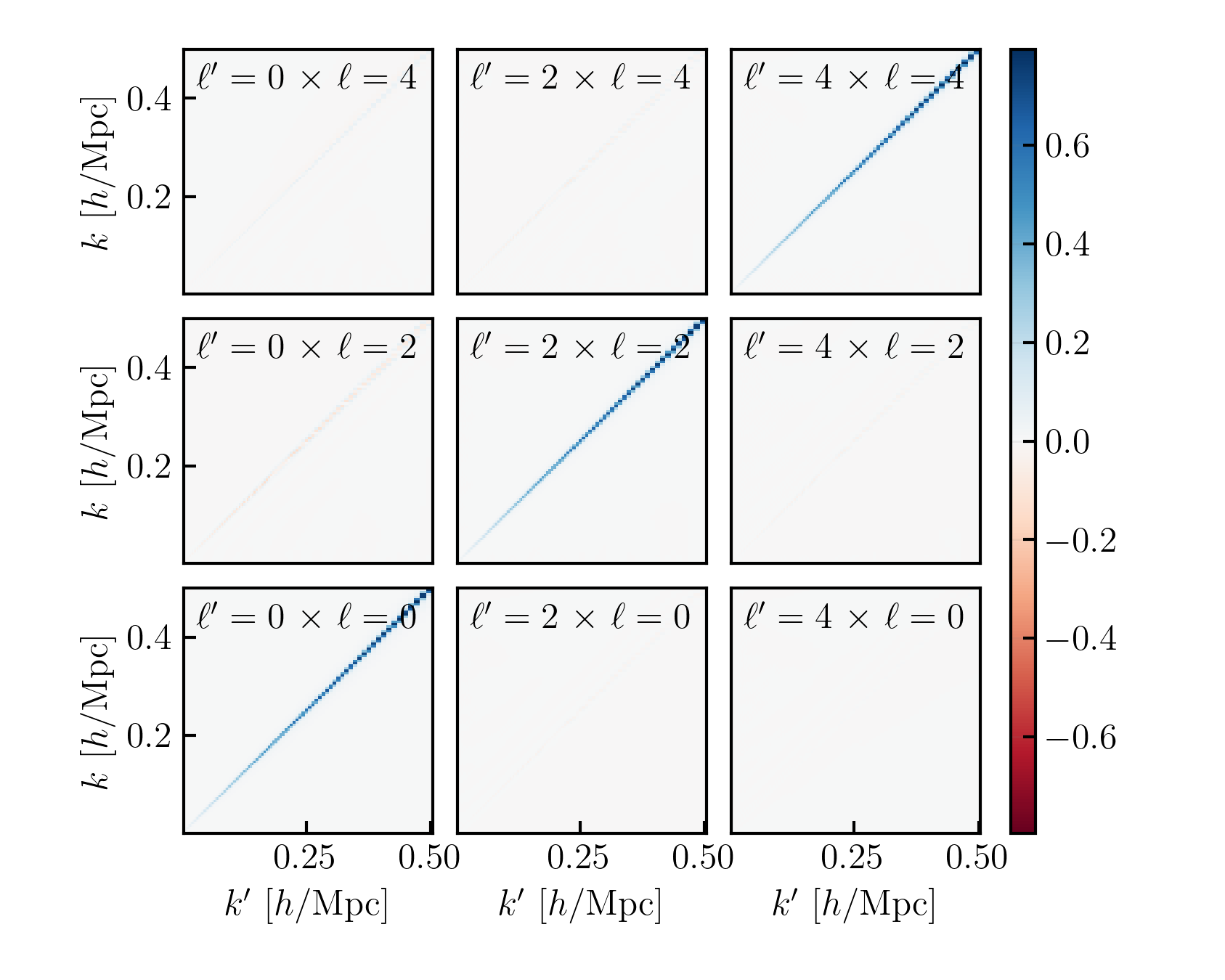}
\end{center}
\caption{Window matrix
without $\theta$-cut 
$W_{\ell \ell^{\prime}}(k, k^{\prime})$ from~\eqref{eq:window_matrix_wa}, computed from the 25 complete ELG AbacusSummit mocks (in $1.1 < z < 1.6$) for $k$, $k^{\prime} < 0.5 \; h/\Mpc$.}
\label{fig:wmatrix}
\end{figure}

\begin{figure}
\begin{center}
\includegraphics[scale=0.8]{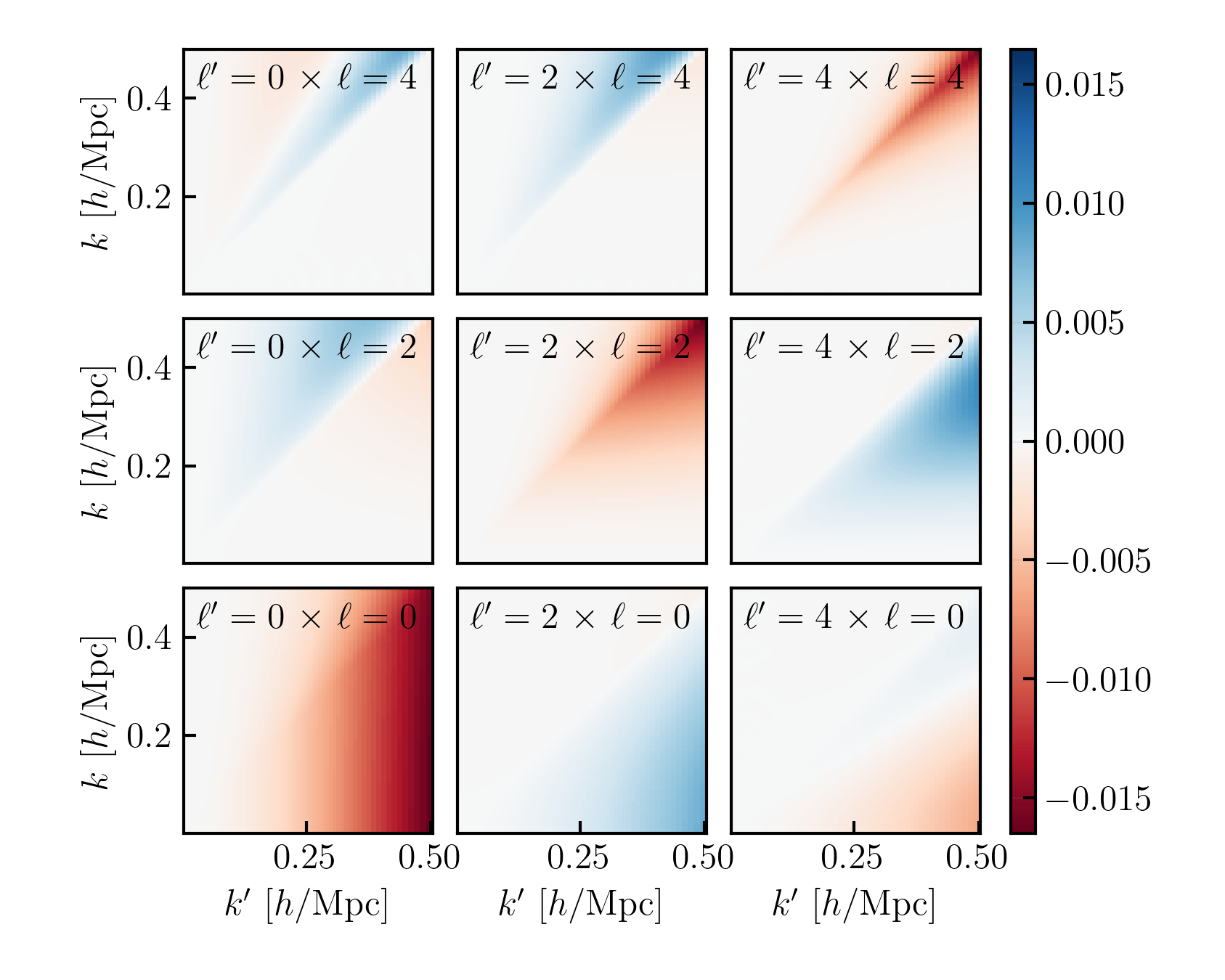}
\end{center}
\caption{Contribution of $\theta < 0.05^{\circ}$ pairs to the window matrix in \eqref{eq:window_matrix_wa},
$\Delta W_{\ell \ell^{\prime}}(k, k^{\prime})$, computed from the 25 complete ELG AbacusSummit mocks (in $1.1 < z < 1.6$), for $k$, $k^{\prime} < 0.5 \; h/\Mpc$.
}
\label{fig:wmatrix_thetacut_diff}
\end{figure}

Figure~\ref{fig:wmatrix} shows the window matrix $W_{\ell \ell^{\prime}}(k, k^{\prime})$ of the merged 25 complete AbacusSummit mocks without $\theta$-cut. We see that the window matrix is very close to diagonal. Figure~\ref{fig:wmatrix_thetacut_diff} shows 
$\Delta W_{\ell \ell^{\prime}}(k, k^{\prime})$, the contribution of galaxy pairs at $\theta < 0.05^{\circ}$ to the complete window matrix,
from the same mocks. $\theta$-cut adds contributions from large theory modes $k^{\prime}$ to all observed modes $k$. 
Note however the scale of the contributions from $\Delta W_{\ell \ell^{\prime}}(k, k^{\prime})$, which are small compared to the diagonal in $W_{\ell \ell^{\prime}}(k, k^{\prime})$ shown in figure~\ref{fig:wmatrix}. In section~\ref{sec:removing_sensitivity_to_high_k_theory}, we present a method to transform the window matrix so that it converges to zero at high $k^\prime$ without modifying the likelihood.

\subsection{Practical computation of the window}\label{sec:window}

The window function, $W_{p}^{(n)}(s)$, written in equation~\eqref{eq:power_window_function}, is computed using \texttt{pypower}. The Fourier space window function (i.e. the Hankel transform of equation~\eqref{eq:power_window_function}) is computed from the random catalog using three meshes with different side lengths, to capture both small and large scales: one equal to that used to compute the power spectrum, i.e. $9 \; \mathrm{Gpc}/h$, one of $45 \; \mathrm{Gpc}/h$ and one of $180 \; \mathrm{Gpc}/h$. The mesh size for the three windows is fixed to that of the power spectrum estimation, i.e. $1500^3$. Each window is computed similarly to the power spectrum in equation~\eqref{sec:power_estimator}, with a fine $k$-binning of $\Delta k = 2\pi/180000 = 3.5 \; \times 10^{-5} \; h/\Mpc$. The three windows are then concatenated together: the $k$-range is first taken from the $180 \; \mathrm{Gpc}/h$ window, and is then 
extended at higher $k$ with the $45 \; \mathrm{Gpc}/h$ and $9 \; \mathrm{Gpc}/h$ windows, such that in each $k$-bin, the window with the larger number of modes in the bin is selected. This ensures that the concatenated window has a high $k$ resolution in a wide range of $k$ values (up to the Nyquist frequency of the smaller box, here $0.52 \; h/\Mpc$). The Fourier space window is then Hankel-transformed to get the configuration space window $W_{p}^{(n)}(s)$ ($n = 0, 1$), which is used to compute the window matrix $W_{\ell \ell^{\prime}}(k, k^{\prime})$ from equation~\eqref{eq:window_matrix_wa}. The contribution from small angular pairs $\Delta W_{p}^{(n)}(s)$ (equation~\eqref{eq:window_direct}) is computed the same way as for the power spectrum estimation, with the same binning in separation $s$ (see appendix~\ref{sec:thetacut_powerspectrum_estimator_numerical_details}). We demonstrate the validity of the window computation in the next section.

\subsection{Validation of the window computation}\label{sec:window_validation}

\begin{figure}
\centering
\includegraphics[scale=0.78]{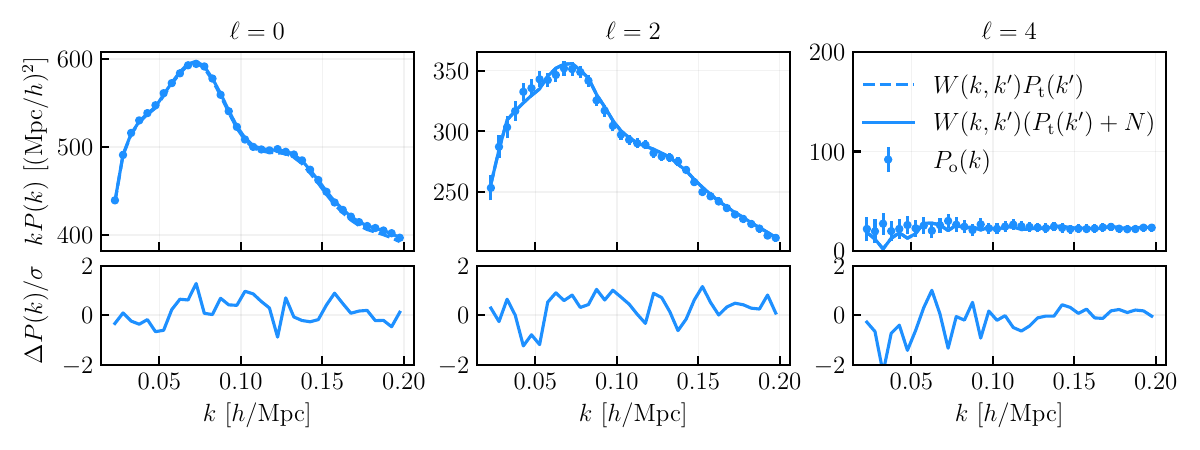}
\caption{
{\it Top:} average power spectrum multipoles ($\ell$ = 0, 2, 4) of the 25 DR1 complete ELG AbacusSummit mocks in the redshift bin $1.1 < z < 1.6$: dots with error bars stand for measurements from cut-sky mocks, solid (resp. dashed) blue lines are predictions from cubic mock power spectra multiplied by the relevant window matrix multipoles without (resp. with) a shot noise term added to the theory. {\it Bottom:} difference between measured and predicted (with shot noise) multipoles, normalized by
1/5th of the standard deviation of the DESI DR1 power spectrum, estimated from 
approximate mocks (see section~\ref{sec:EZmocks_covariance}).
}
\label{fig:window_validation_complete_nocut}
\end{figure}

\begin{figure}
\centering
\includegraphics[scale=0.78]{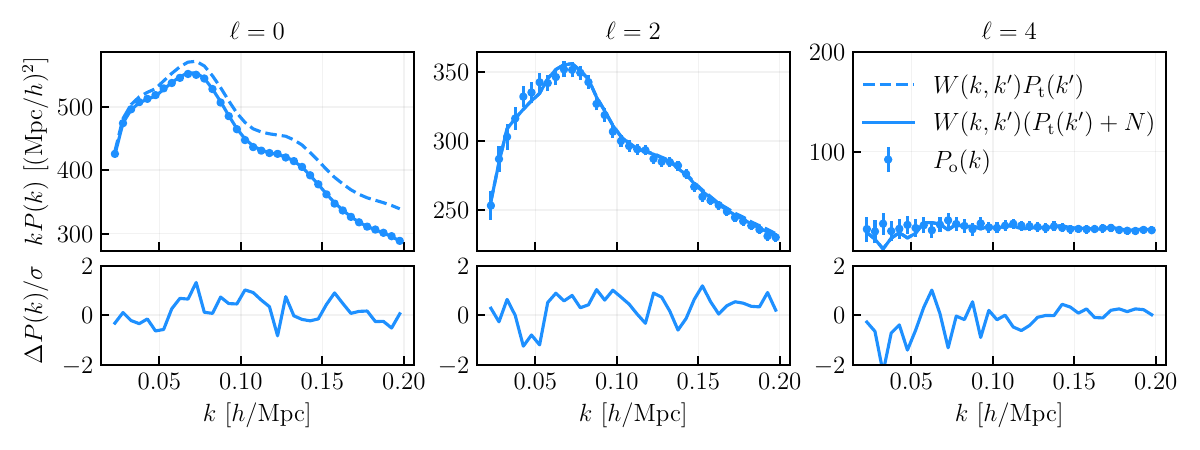}
\caption{Same as figure~\ref{fig:window_validation_complete_nocut} with $\theta$-cut applied.  }
\label{fig:window_validation_complete_thetacut}
\end{figure}

In this section we demonstrate the validity of the power spectrum window matrix computation. In order to approximate the underlying theory $\Pt$ of the mocks, we use the average power spectrum of the 25 cubic AbacusSummit simulations from which the ELG $1.1 < z < 1.6$ mocks were created. Cubic power spectrum multipoles in the $k^{\prime}$ range $[0.001, 0.35] \; h/\Mpc$ are used as theory, and multiplied by the window matrix to get a prediction for the multipoles that will be measured from the cut-sky mocks. 
Figure~\ref{fig:window_validation_complete_nocut} shows the comparison between the average power spectrum multipoles of the 25 complete cut-sky mocks, $\Po$, and the average of the power spectrum multipoles from the cubic mocks multiplied by the corresponding window matrix multipoles $W_{\ell \ell^{\prime}}(k, k^{\prime})$. Additionally, since a constant shot noise term added to the theory is usually left free in the fits (see section~\ref{sec:full_shape_fits_power}), we add a free constant term $N$ to the theoretical prediction for the monopole, before the product with the window matrix. We fit $N$ to minimize residuals between the theoretical prediction from the cubic mocks and the average power spectrum multipoles of the cut-sky mocks. We find that allowing this shot noise term $N$ to vary gives a slightly better agreement between the windowed theory and cut-sky power spectrum. Indeed, the cut-sky mocks are downsampled with respect to cubic boxes to get the right average number density, therefore we cannot use the with-shot-noise cube power as a ground-truth
theory for the cut-sky mocks. Moreover, this shot noise term absorbs the residual errors due to the resolution of the mesh used to compute the window and the truncation of the theory modes at $k^{\prime} < 0.35 \; h/\Mpc$ (especially in the $\theta$-cut case). The bottom panels of Figure~\ref{fig:window_validation_complete_nocut} show the difference between $\Po(k)$ and $W_{\ell \ell^{\prime}}(k, k^{\prime})(\Pt(k^{\prime})+N)$, normalized by 1/5th of the standard deviation estimated from approximate mocks (see section~\ref{sec:EZmocks_covariance}). 
The cut-sky power spectrum multipoles and the cubic power spectrum multipoles multiplied by the window matrix are in good agreement, within 1/5th of DR1 statistical error. Figure~\ref{fig:window_validation_complete_thetacut} is the same as figure~\ref{fig:window_validation_complete_nocut} with $\theta$-cut applied in both $\Po(k)$ and the window $W_{\ell \ell^{\prime}}(k, k^{\prime})$. Note that the residuals in figure~\ref{fig:window_validation_complete_thetacut} are almost identical to those of figure~\ref{fig:window_validation_complete_nocut}. Indeed, the fitting for $N$ compensates for the high-$k^{\prime}$ truncation of the theory (see the discussions in sections~\ref{sec:full_shape_fits_power} and~\ref{sec:removing_sensitivity_to_high_k_theory}).

\section{Application to full-shape analysis}\label{sec:full_shape_fits}
In this section we test the ability of our method to yield unbiased cosmological constraints from fiber-assigned mocks by doing a full-shape fit --- as opposed to Baryon Acoustic Oscillation (BAO) analyses that only extract the BAO information from two-point measurements --- of the correlation function and power spectrum multipoles with $\theta$-cut. We sample cosmological parameters posterior distributions from the average correlation function and power spectrum multipoles of the 25 AbacusSummit ELG cut-sky mocks, as well as from each individual mock, in four configurations: complete mocks, complete mocks with $\theta$-cut, \altmtl\ mocks and \altmtl\ mocks with $\theta$-cut. We show that, for complete mocks, the $\theta$-cut by itself introduces a negligible bias in cosmological parameter estimates with respect to DR1 statistical uncertainty. Moreover, although constraints on cosmological parameters from \altmtl\ mocks are biased with respect to that from complete mocks when no correction is applied, we show that the $\theta$-cut allows to reduce this bias.  

\subsection{Mock-based covariance matrix}\label{sec:EZmocks_covariance}
To generate the covariance matrix of the data, we use 1000 approximate simulations based on the Zel'dovich approximation~\cite{zeldovich_gravitational_1970}, called EZmocks, that have clustering properties matching $N$-body simulations in terms of one-point, two-point and three-point statistics~\cite{chuang_ezmocks_2015}. The set of EZmocks we use are built from a cubic snapshot at $z = 1.325$ simulating ELG clustering, and sampled to match the footprint and redshift density distribution $n(z)$ of the DESI DR1 ELG sample. In addition, because the \altmtl\ procedure would be too computationnally expensive to run on the 1000 EZmocks, EZmocks are processed through a fast fiber assignment algorithm~\cite{KP3s11-Sikandar}, that emulates the fiber assignment process based on shallow learning.

The power spectrum of each EZmock is computed in 0.005 $h$/Mpc wide $k$-bins in the range $k \in [0, 0.4] \; h/\Mpc$, with 15 times as many randoms as data, otherwise with the same settings as for the 25 AbacusSummit mocks in section~\ref{sec:power_estimator}.
Power spectrum estimates are made with and without $\theta$-cut.
We apply the Hartlap correction~\cite{hartlap_why_2007} to account for the finite number of mocks used to estimate the covariance matrix. The correlation function covariance matrix is obtained similarly.

\subsection{Fitting setup}\label{sec:fit_setup}
We fit the correlation function and power spectrum of the ELG cut-sky mocks with the Lagrangian Pertubation Theory model \texttt{velocileptors} \cite{chen_consistent_2020}, and with ShapeFit~\cite{brieden_shapefit_2021} template, as implemented in \texttt{desilike}.\footnote{\url{https://github.com/cosmodesi/desilike}.} ShapeFit template includes parameters for the Alcock-Paczynski effect, a power spectrum tilt parameter and the growth rate. Our parametrization of the Alcock-Paczynski effect is defined by $q_{\mathrm{iso}}$ and $q_{\mathrm{ap}}$ as follows:
\begin{align}
    q_{\mathrm{iso}} = q_{\parallel}^{\frac{1}{3}} q_{\perp}^{\frac{2}{3}}, \;
    q_{\mathrm{ap}} = \frac{q_{\parallel}}{q_{\perp}}
\end{align}
where $q_\parallel$ and $q_\perp$ are scalings along and transverse to the line-of-sight, respectively, and are defined through:
\begin{align}
    P^{\mathrm{obs}}(\boldsymbol{k}^{\mathrm{obs}}) = q_\perp^{-2} q_\parallel^{-1} P(\boldsymbol{k}), \;
    k^{\mathrm{obs}}_{\parallel, \perp} =  q_{\parallel, \perp}\  k_{\parallel, \perp}.
\end{align}
where $P(\boldsymbol{k})$ is the theoretical power spectrum in terms of the true coordinate $\boldsymbol{k}$ and $P^{\mathrm{obs}}(\boldsymbol{k}^{\mathrm{obs}})$ in terms of the observed coordinate $\boldsymbol{k}^{\mathrm{obs}}$. Additionally, we parametrize the growth rate in terms of its value with respect to the fiducial cosmology:
\begin{equation}
df = \frac{f}{f^{\mathrm{fid}}},
\end{equation}
where the superscript $^{\mathrm{fid}}$ denotes the value in the fiducial cosmology, while $dm$ parametrizes the change in the slope of the power spectrum at the pivot scale $k_p$~\cite{brieden_shapefit_2021}:
\begin{align}
    P^{\prime}_{\mathrm{lin}}(k) = P_{\mathrm{lin}}^{\mathrm{fid}}(k)\ \exp\left\{ \frac{dm}{a}\tanh \left[a\ln\left(\frac{k}{k_p}\right) \right] \right\}, 
    \label{eq: plin_sf}
\end{align}
with $a = 0.6$ and $k_p = 0.03\; h/\Mpc$. Other parameters of the model, defined in~\cite{brieden_shapefit_2021}, include linear, second-order and third-order bias parameters $b_1$, $b_2$, $b_3$, tidal bias parameter $b_s$, stochastic parameters for each multipole $s_{n, 0}$, $s_{n, 2}$, $s_{n, 4}$ and coefficients of Effective Field Theory counterterms of the form $\mu^{n}$: $\alpha_{0}$, $\alpha_{2}$, $\alpha_{4}$, $\alpha_{6}$, the latter of which is fixed. $b_3$ is fixed to 0.

Our fiducial cosmology is AbacusSummit base cosmology, which is $\Lambda$CDM cosmology from Planck 2018~\cite{planck_collaboration_planck_2020}, as summarized in table~\ref{tab:fiducial_cosmology}. Table~\ref{tab:shapefit_priors} shows the priors for all varied parameters.
\begin{table*}
\centering
\begin{tabular}{c|c|c}
\hline
 & description & fiducial value \\
\hline
$\omega_{\mathrm{cdm}}$ & density of cold dark matter & 0.1200 \\
$\omega_{\mathrm{b}}$ & baryon density & 0.02237 \\
$h$ & reduced Hubble parameter & 0.6736 \\
$A_s$ & curvature power spectrum value at $k_{\mathrm{pivot}}$ & $2.0830 \times 10^{-9}$ \\
$n_s$ & scalar spectral index & 0.9649 \\
\hline
\end{tabular}
\caption{AbacusSummit base cosmology from Planck2018~\cite{planck_collaboration_planck_2020} used as the fiducial cosmology in this analysis. $k_{\mathrm{pivot}}$ is fixed to 0.05 $h$/Mpc.}
\label{tab:fiducial_cosmology}
\end{table*}

\begin{table*}
\centering
\begin{tabular}{c|c}
\hline
parameter & prior \\
\hline
$q_{\mathrm{iso}},\; q_{\mathrm{ap}}$ & $\mathcal{U}[0.8, 1.2]$ \\
$df$ & $\mathcal{U}[0, 2]$ \\
$dm$ & $\mathcal{U}[-3, 3]$ \\
$b_1$ & $\mathcal{U}[0, 3]$ \\
$b_2,\; b_s$ & $\mathcal{N}(0, 5)$ \\
$s_{n, 0},\; s_{n, 2}\;, s_{n, 4}$ & $\mathcal{N}(0, 10)$  \\
$\alpha_{0},\; \alpha_{2}\;, \alpha_{4}$ & $\mathcal{N}(0, 10)$  \\
\hline
\end{tabular}
\caption{Priors on each varied parameter. $\mathcal{U}[{\rm min, max}]$ refers to a uniform distribution within $[{\rm min, max}]$ and $\mathcal{N}(\mu,\sigma)$ to a normal distribution with mean $\mu$ and standard deviation $\sigma$.}
\label{tab:shapefit_priors}
\end{table*}

Our data vector $\vd$ consists in the concatenation of the measured (average or from individual mocks) power spectrum multipoles of order $\ell$ = 0, 2, 4 (monopole, quadrupole and hexadecapole) in the $k$-range $0.02 < k \; [h/\Mpc] < 0.2$ with $0.005 \; h/\Mpc$ spacing. We relate the theoretical multipoles $\vt$ to the observed multipoles through the window matrix defined in section~\ref{sec:power_window}. As a result, given the covariance matrix $\mathrm{C}$, the likelihood reads:
\begin{equation}
    \mathcal{L}(\vd | \vt) = \exp \left( {- \frac{1}{2} \left( \vd - \mathrm{W} \vt \right)^{T} \mathrm{C}^{-1} \left( \vd - \mathrm{W} \vt \right)} \right).
\end{equation}
The likelihood for the correlation function is defined similarly, where $\vd$ stands for the concatenated correlation function multipoles measured from the mocks in the separation range $30 < s \; [\Mpc/h] < 150$ with $4 \; \Mpc/h$ spacing, $\vt$ the theoretical correlation function and W the window matrix defined in~\eqref{eq:window_cut_correlation}. The same model and template as for the power spectrum is used. The posterior is sampled with Monte Carlo Markov Chains (MCMC) with \texttt{emcee}~\cite{foreman-mackey_emcee_2013}.\footnote{\url{https://github.com/dfm/emcee}.}

\subsection{Full-shape fits to \texorpdfstring{$\theta$}{theta}-cut correlation function}

We fit the average correlation function of the 25 AbacusSummit simulations. Moreover, in order to estimate the errors on the bias in parameter constraints induced by the $\theta$-cut, we want to obtain posterior distributions from each individual mock. Since the posteriors for the average of complete mocks with or without $\theta$-cut are very close to each other, to minimize sampling noise, instead of re-sampling MCMC from each mock, we perform importance sampling~\cite{kloek_bayesian_1978} from the chains run on each uncut complete mock to get the posteriors for $\theta$-cut complete mocks. Namely, to estimate the parameters' posterior distribution for a given $\theta$-cut complete mock we weight each of the samples of the corresponding uncut complete mock chains by the ratio of the $\theta$-cut to uncut complete mock posteriors. For \altmtl\ mocks we do not resort to importance sampling, since mock posteriors with and without $\theta$-cut differ significantly, and we rather sample them independently.

\begin{figure}
\centering
\includegraphics[scale=0.8]{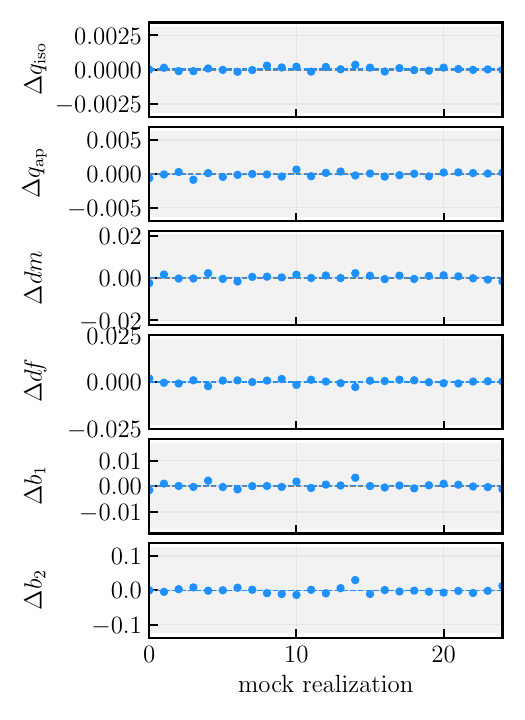}
\includegraphics[scale=0.8]{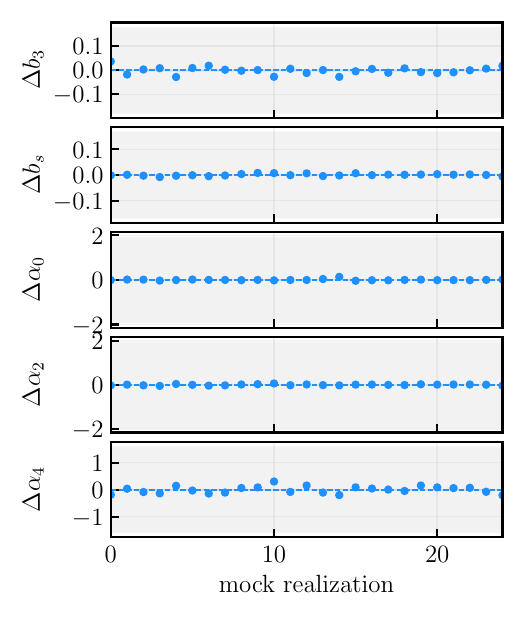}
\caption{Difference between correlation function fits with and without $\theta$-cut for each varied parameter. Blue dots indicate the difference in the mean posterior value with and without $\theta$-cut for each complete mock. The horizontal blue dashed line shows the mean of the difference across the 25 mocks. The grey area (filling the entire panels) indicates 1/5th of the DR1 statistical uncertainty on each parameter, obtained by sampling posteriors with the average $\theta$-cut correlation function of the 25 \altmtl\ mocks.
}
\label{fig:corr_mocks_importancesampling_thetacut_dispersion}
\end{figure}

First, let us verify that $\theta$-cut itself does not bias the estimates of cosmological parameters, using complete mocks. Figure~\ref{fig:corr_mocks_importancesampling_thetacut_dispersion} shows the deviation of the mean posterior value of the fit with $\theta$-cut with respect to that of the fit without $\theta$-cut, for each of the free parameters, and for each of the 25 complete mocks. The grey area (filling most of each panel) represents 1/5th of DR1 statistical error on each parameter, as given by the posterior errors obtained with the average correlation function of the 25 \altmtl\ mocks with $\theta$-cut, using EZmocks covariance matrix (with $\theta$-cut). The horizontal blue dashed line indicates the average shift over the 25 mocks. The uncertainty on this shift is at most $0.2\%$ of DR1 uncertainty for $q_{\mathrm{iso}}, q_{\mathrm{ap}}, df, dm$, which indicates that 25 mocks are largely sufficient to estimate the bias induced by the $\theta$-cut. As reported in the first column of table~\ref{tab:corr_thetacut_bias}, the $\theta$-cut by itself biases constraints on cosmological parameters by at most $0.1 \%$ of DR1 statistical uncertainty.

\begin{figure}
\centering
\includegraphics[scale=0.78]{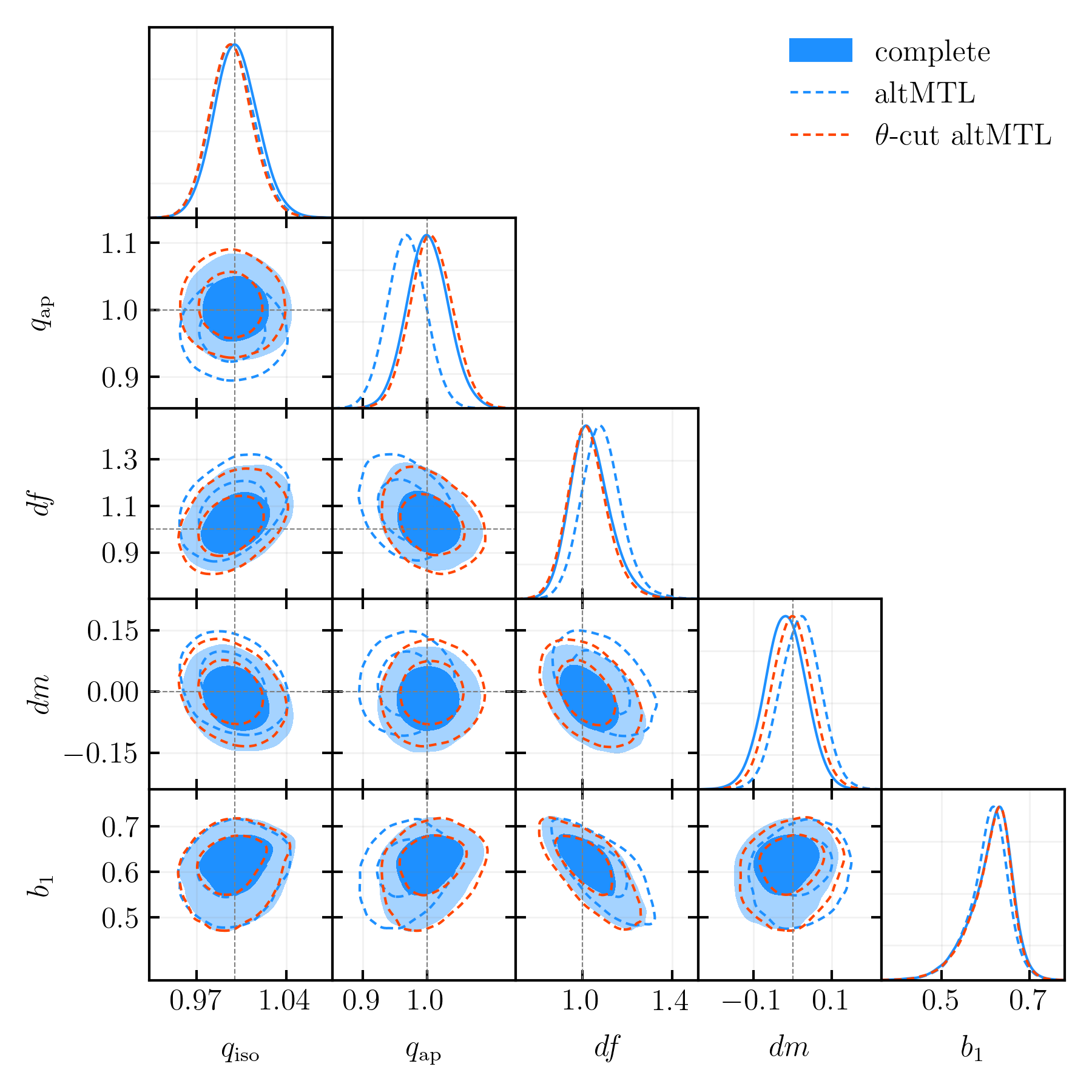}
\caption{Marginalized posteriors on cosmological parameters $q_{\mathrm{iso}}$, $q_{\mathrm{ap}}$, $df$, $dm$ and linear bias $b_1$, obtained from the average correlation function multipoles ($\ell = 0, 2, 4$) of the 25 AbacusSummit mocks. Different samples have been analyzed: complete samples (blue), \altmtl\ samples (dashed blue) and \altmtl\ samples with $\theta$-cut applied (dashed red). Inner and outer contours enclose $68\%$ and $95\%$ of the projected 2D distribution, respectively. Dashed lines represent fiducial parameter values compared to which posteriors for complete mocks are slightly shifted (e.g. $dm$) due to the intrinsic variance on the mean of the 25 mocks.}
\label{fig:corner_plot_corr_thetacut}
\end{figure}

Now let us assess the ability of $\theta$-cut to unbias the constraints on cosmological parameters obtained from \altmtl\ mocks. Figure~\ref{fig:corner_plot_corr_thetacut} shows the posterior distributions for the average correlation function of the 25 AbacusSummit mocks. Filled blue contours show the posteriors for complete mocks, dashed blue contours for \altmtl\ mocks and dashed red contours for $\theta$-cut \altmtl\ mocks. In all cases we use the covariance matrix from EZmocks, with or without $\theta$-cut in the corresponding cases. The width of the posteriors with $\theta$-cut, which are not shown in this figure, are virtually unchanged with respect to that of the uncut posteriors. Constraints on $q_{\mathrm{ap}}$, $dm$ and $df$
for \altmtl\ mocks are biased at the level of 100\%, 77\% and 52\% of DR1 statistical uncertainty compared to complete mocks, respectively. Applying the $\theta$-cut to \altmtl\ mocks allows these biases to be reduced down to 22\%, 33\% and 14\% of the DR1 uncertainty, respectively (see table~\ref{tab:corr_thetacut_bias}). The constraint on $q_{\mathrm{iso}}$ is not affected by the $\theta$-cut, and stays within 18\% of DR1 uncertainty.
We note however that these numbers are intrinsically limited by the number of mocks available and show no detection of a remaining bias. We indeed compute the uncertainty on the mean over the 25 mocks \altmtl-to-complete mock dispersion to be of 12\% of DR1 uncertainty for $q_{\mathrm{iso}}$, 19\% of the uncertainty for $q_{\mathrm{ap}}$, 15\% for $dm$ and 18\% for $df$.

\begin{table*}
\centering
\begin{tabular}{c|cccc}
\hline
& $\theta$-cut complete   & altMTL & $\theta$-cut altMTL &   DR1 error \\
\hline
 $q_{\mathrm{iso}}$ $-$ $q_{\mathrm{iso}}$$(\mathrm{complete})$ & $0.00001 \pm 0.00003$   & $-0.003 \pm 0.002$ & $-0.003 \pm 0.002$    &       0.017 \\
 $q_{\mathrm{ap}}$ $-$ $q_{\mathrm{ap}}$$(\mathrm{complete})$   & $-0.00002 \pm 0.00007$  & $-0.032 \pm 0.006$ & $0.007 \pm 0.006$     &       0.032 \\
 $dm$ $-$ $dm$$(\mathrm{complete})$                             & $0.00007 \pm 0.00007$   & $0.040 \pm 0.008$  & $0.017 \pm 0.008$     &       0.052 \\
 $df$ $-$ $df$$(\mathrm{complete})$                             & $0.00012 \pm 0.00018$   & $0.047 \pm 0.016$  & $-0.013 \pm 0.016$    &       0.091 \\
\hline
\end{tabular}
\caption{Shift in the mean posterior values obtained with $\theta$-cut complete, \altmtl\ and $\theta$-cut \altmtl\ correlation function mocks with respect to uncut complete mocks. The third column (DR1 error) is the standard deviation of the sample from the mean correlation function of the 25 $\theta$-cut \altmtl\ AbacusSummit mocks.}
\label{tab:corr_thetacut_bias}
\end{table*}

\subsection{Full-shape fits to \texorpdfstring{$\theta$}{theta}-cut power spectrum}\label{sec:full_shape_fits_power}

\begin{figure}
\centering
\includegraphics[scale=0.8]{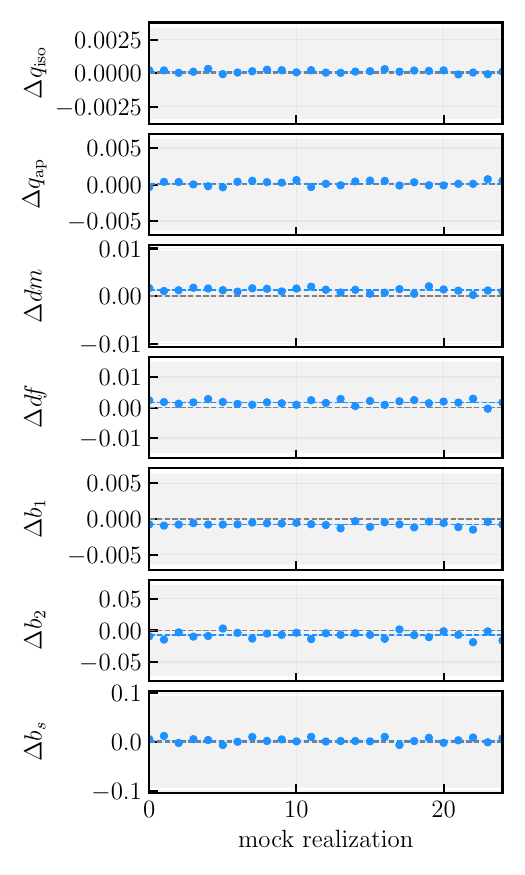}
\includegraphics[scale=0.8]{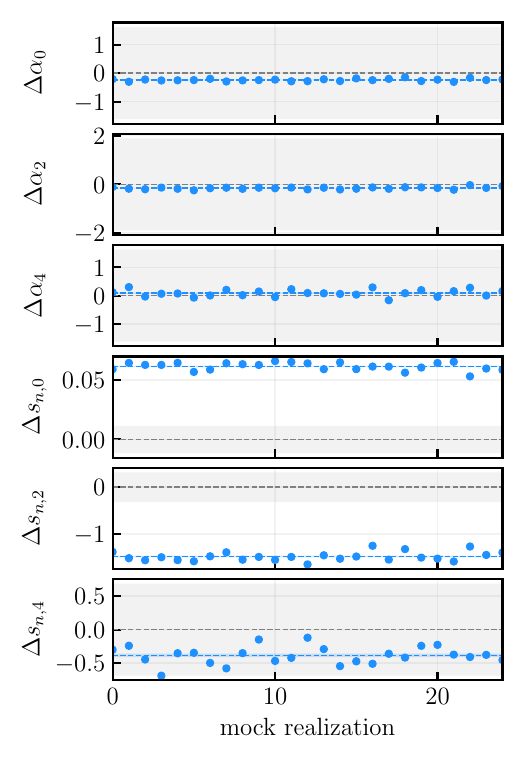}
\caption{Difference of constraints obtained from power spectrum samples with and without $\theta$-cut for each varied parameter. Blue dots indicate the difference in the mean posterior value with and without $\theta$-cut for each complete mock. The horizontal blue dashed line shows the mean of the difference across the 25 mocks. The grey area (filling most panels entirely) indicates 1/5th of DR1 statistical uncertainty on each parameter, obtained by sampling posteriors with the average $\theta$-cut power spectrum of the 25 \altmtl\ mocks.}
\label{fig:power_mocks_importancesampling_thetacut_dispersion}
\end{figure}

Figure~\ref{fig:power_mocks_importancesampling_thetacut_dispersion} shows the deviation of the mean posterior value of the fit with $\theta$-cut compared to the fit without $\theta$-cut for all free parameters. As reported in table~\ref{tab:power_thetacut_bias}, the uncertainty on the mean shift between $\theta$-cut complete and complete mocks is at most $0.2 \%$ of DR1 uncertainty. The $\theta$-cut by itself biases constraints on cosmological parameters to
at most 2.5\% of DR1 uncertainty. As shown in figure~\ref{fig:power_mocks_importancesampling_thetacut_dispersion}, the stochastic parameters $s_{n, 0}$ and $s_{n, 2}$ are also significantly shifted by the $\theta$-cut, at the level of $\sim$ 107\% and $\sim$ 94\% of DR1 statistical uncertainty, respectively. This can be explained by the fact that these parameters absorb the loss of the contribution from the high theory-$k$ values of the $\theta$-cut window matrix (see figure~\ref{fig:wmatrix_thetacut_diff}) due to the theory cut $k^\prime < 0.35 \; \mathrm{Mpc}/h$. Indeed, as the $\theta$-cut window couples information from the small-scale theory in the observed large-scale modes (see figure~\ref{fig:wmatrix_thetacut_diff}), when we truncate the theory to scales $k^\prime < 0.35 \; \mathrm{Mpc}/h$, even large scales can be incorrectly modelled. The change in the values of $s_{n, 0}$ and $s_{n, 2}$ compensates this error.

\begin{figure}
\centering
\includegraphics[scale=0.78]{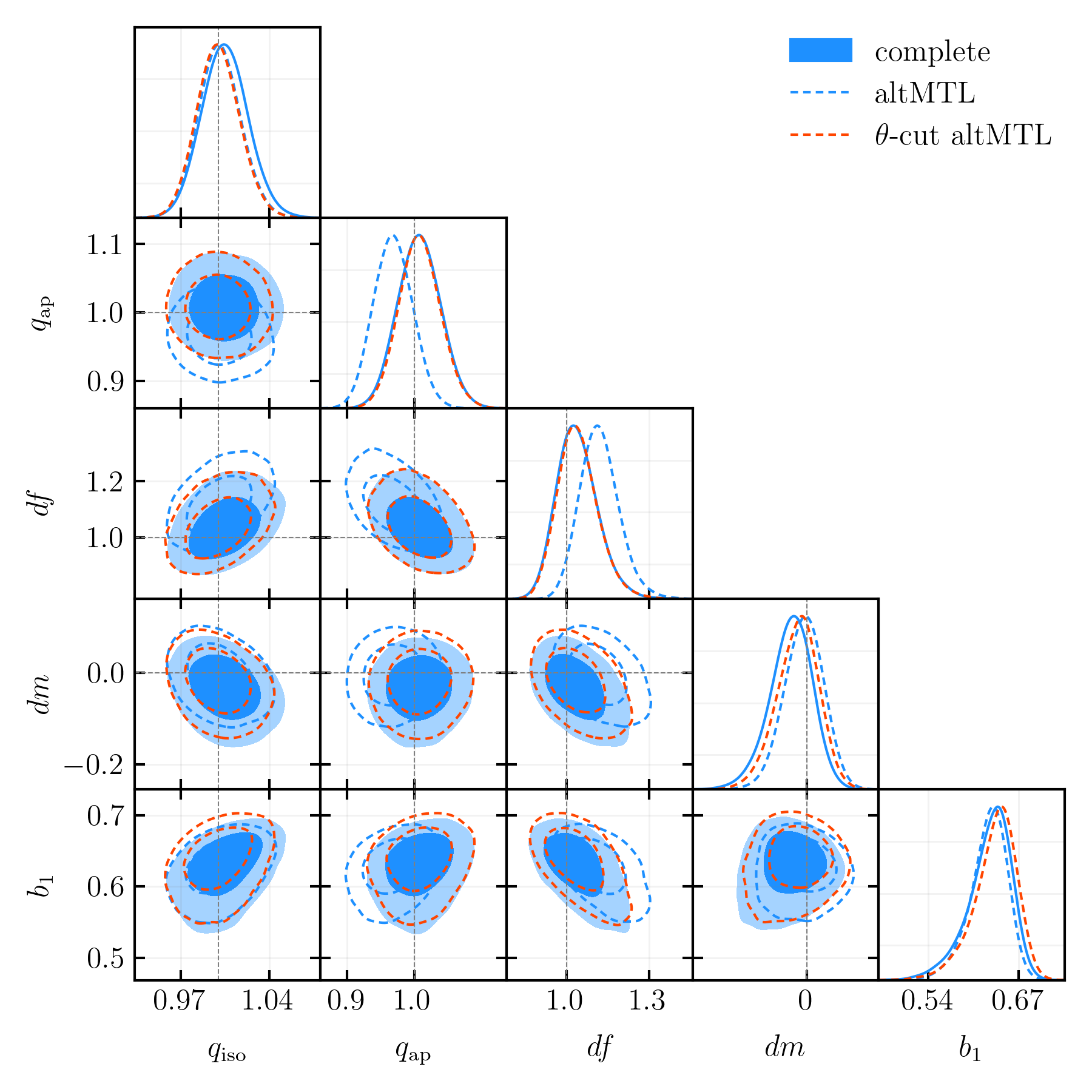}
\caption{Marginalized posteriors on cosmological parameters $q_{\mathrm{iso}}$, $q_{\mathrm{ap}}$, $df$, $dm$ and linear bias $b_1$, obtained from the average power spectrum multipoles ($\ell = 0, 2, 4$) of the 25 AbacusSummit mocks. Different samples have been analyzed: complete samples (blue), \altmtl\ samples (dashed blue) and \altmtl\ samples with $\theta$-cut applied (dashed red).}
\label{fig:corner_plot_power_thetacut}
\end{figure}

Figure~\ref{fig:corner_plot_power_thetacut} shows that fiber assignment incompleteness biases constraints on $q_{\mathrm{ap}}$, $dm$ and $df$ from power spectrum measurements to 118\%, 58\% and 104\% of the DR1 uncertainty, respectively (see table~\ref{tab:power_thetacut_bias}). For these parameters, the $\theta$-cut allows the same constraints to be recovered as with complete mocks within
13\%, 26\% and 3\% of the DR1 uncertainty, respectively. The $q_{\mathrm{iso}}$ parameter is the only one for which the bias is not reduced by the $\theta$-cut, but remains at the level of 28\% of DR1 uncertainty. As for the correlation function, these numbers are intrinsically limited by the number of mocks available and show no detection of a remaining bias. We indeed compute the uncertainty on the mean over the 25 mocks due to \altmtl-to-complete mock dispersion to be of 18\% of DR1 uncertainty for $q_{\mathrm{iso}}$, 19\% of the uncertainty for $q_{\mathrm{ap}}$, 18\% for $dm$ and 16\% for $df$.

\begin{table*}
\centering
\begin{tabular}{c|cccc}
\hline
& $\theta$-cut complete   & altMTL & $\theta$-cut altMTL &   DR1 error \\
\hline
$q_{\mathrm{iso}}$ $-$ $q_{\mathrm{iso}}$$(\mathrm{complete})$ & $(12 \pm 2) \times 10^{-5}$   & $-0.005 \pm 0.003$ & $-0.005 \pm 0.003$  &       0.017 \\
 $q_{\mathrm{ap}}$ $-$ $q_{\mathrm{ap}}$$(\mathrm{complete})$   & $(19 \pm 6) \times 10^{-5}$   & $-0.037 \pm 0.006$ & $0.004 \pm 0.006$   &       0.031 \\
 $dm$ $-$ $dm$$(\mathrm{complete})$                             & $0.0012 \pm 0.0001$    & $0.028 \pm 0.008$  & $0.013 \pm 0.008$   &       0.048 \\
 $df$ $-$ $df$$(\mathrm{complete})$                             & $0.0017 \pm 0.0002$    & $0.078 \pm 0.011$  & $0.002 \pm 0.012$   &       0.075 \\
\hline
\end{tabular}
\caption{Shift in the mean posterior values obtained with $\theta$-cut complete, \altmtl\ and $\theta$-cut \altmtl\ power spectrum mocks with respect to uncut complete mocks. The third column (DR1 error) is the standard deviation of the sample from the mean power spectrum of the 25 $\theta$-cut \altmtl\ AbacusSummit mocks.}
\label{tab:power_thetacut_bias}
\end{table*}

\section{Removing sensitivity to high theory \texorpdfstring{$k$}{k} modes}
\label{sec:removing_sensitivity_to_high_k_theory}

We showed in section~\ref{sec:full_shape_fits_power} that using the model described in section~\ref{sec:fit_setup}, fits with $\theta$-cut power spectra give unbiased constraints on cosmological parameters, suggesting that non-zero terms at high theory modes $\kt$ in $\mathrm{W} \vt$ (coming from the non-zero columns at high $\kt$ in $\mathrm{W}$), truncated by our theory cut $\kt < 0.35 \; \mathrm{Mpc}/h$, are absorbed by stochastic parameters. While these terms are apparently not problematic in our specific case, in general we do not want the window matrix to have non-zero contributions from the theory at $\kt$ values above our fitting range, that is the maximum $\kt$ where the theory is reliable. An idea to avoid having diverging tails at high $\kt$ would be to apply a smooth cut instead of the sharp truncation we have been doing. This idea of apodization is explored in appendix~\ref{sec:apodization}, where we show that we would need to remove galaxy pairs at transverse separations of up to $\sim 30\; \Mpc/h$ to get a vanishing window at $\kt \sim 0.3-0.4 \; h/\mathrm{Mpc}$. In this section, we apply an alternative method developed in~\cite{mcdonald_window} that consists in performing a change of basis that forces the non-zero elements of the window to be concentrated in a $\kt$ range as narrow as possible, while leaving the likelihood unchanged. 

\subsection{Transforming the window matrix to make it more compact in \texorpdfstring{$\kt$}{kt}}

In this subsection, we describe our method to find a transformation matrix $\mathrm{M}$ to apply to the window matrix $\mathrm{W}$ such that $\mathrm{M}\mathrm{W}$ converges towards zero at high theory modes $\kt$, while conserving the likelihood of the data. The log-likelihood is proportional to the $\chi^2$:
\begin{equation}
    \chi^2 = \left( \vd - \mathrm{W} \vt \right)^T \mathrm{C}^{-1} \left( \vd - \mathrm{W} \vt \right)
\end{equation}
where $\vd$ is the data vector, $\vt$ is the theory, and $\mathrm{C}$ the covariance matrix. Provided that M is invertible, we can write:
\begin{equation}
    \chi^2 = \left( \mathrm{M} \vd - \mathrm{M} \mathrm{W} \vt \right)^T (\mathrm{M}\mathrm{C}\mathrm{M}^T)^{-1} \left(  \mathrm{M} \vd -  \mathrm{M} \mathrm{W} \vt \right).
\label{eq:rotated_chi2}
\end{equation}
We found that when applying only the rotation M, removing the contribution of high $\kt$ modes in the window came at the cost of highly distorting the data vector, because the sum over the $\theta$-cut window matrix columns (i.e. the integral of the window matrix over $\kt$) is close to zero. This motivated the addition of a corrective term to allow the window columns to have a non-zero sum. Indeed, we can modify the window matrix by: $\mathrm{W} \xrightarrow{} \mathrm{W} - \vmo \vmt^T$, noting that for any theory $\vt$ and any vector $\vmt$, $\vmt^T \vt$ is a scalar. This means that, provided that we marginalize over $\vmo$ in the data vector, i.e. if we add a term $s \vmo$ with $s$ free, we are removing sensitivity to $\vmo$ in the fit, and thus the additional $\vmo \vmt^T$ term in the window matrix has no impact on our results. So the $\chi^2$ is now:
\begin{equation}
    \chi^2 = \left( \vd^{\prime} - \mathrm{W}^{\prime} \vt \right)^T \mathrm{C}^{\prime \; -1} \left( \vd^{\prime} - \mathrm{W}^{\prime} \vt \right)
\end{equation}
with the transformed data vector, window matrix, and covariance matrix:
\begin{equation}
    \vd^{\prime} = \mathrm{M} \vd - s \vmo,
\end{equation}
\begin{equation}
    \mathrm{W}^{\prime} = \mathrm{M} \mathrm{W} -\vmo \vmt^T,
\end{equation}
\begin{equation}
    \mathrm{C}^{\prime} = \mathrm{M} \mathrm{C} \mathrm{M}^T,
\end{equation}
with $s$ a free parameter.

We want to find $\mathrm{M}$, $\vmo$ and $\vmt$ such that $\mathrm{W}^{\prime}$ is as compact as possible in \ko, and $\mathrm{C}^{\prime}$ as diagonal as possible. We define a loss function to minimize:
\begin{equation}
    L(\mathrm{M}, \vmo, \vmt) = L_{\mathrm{W}}(\mathrm{M}, \vmo, \vmt) + L_{\mathrm{C}}(\mathrm{M}, \vmo, \vmt) + L_{\mathrm{M}}(\mathrm{M})
\end{equation}
with:
\begin{equation}
    L_{\mathrm{W}}(\mathrm{M}, \vmo, \vmt) =  \frac{\sum_{ij} w_{ij} \abs{\mathrm{W}^{\prime}_{ij}}}{\sum_{ij} \abs{\mathrm{W}^{\prime}_{ij}}},
\end{equation}
where $i$ and $j$ indices are spanning the different values of $\ell_{\mathrm{o}}$, $\ko$ and $\ell_{\mathrm{t}}$, $\kt$ respectively, and 
\begin{equation}
    w_{ij} =  \min \left\lbrace \left( \frac{k_{\mathrm{o}, i} - k_{\mathrm{t}, j}}{\sigma_k} \right)^2, \alpha \right\rbrace + \beta \delta_{\ell_{i}\ell_{j}},
\end{equation}
where we chose $\alpha = 5$ and $\beta = 10$, and  $\sigma_k$ is defined to be the width at half maximum of the window peak at $k_{\mathrm{o}} = 0.1 \; h/\Mpc$. $L_{\mathrm{C}}$ is defined as:
\begin{equation}
    L_{\mathrm{C}}(\mathrm{M}, \vmo, \vmt) =  \frac{\sum_{ij} r_{ij} \abs{\mathrm{R}^{\prime}_{ij}}}{\sum_{ij} \abs{\mathrm{R}^{\prime}_{ij}}},
\end{equation}
where $\mathrm{R}^{\prime}$ is the correlation matrix:
\begin{equation}
    \mathrm{R}^{\prime}_{ij} = \frac{\mathrm{C}^{\prime}_{ij}}{\sqrt{\mathrm{C}^{\prime}_{ii} \mathrm{C}^{\prime}_{jj}}},
\end{equation}
and:
\begin{equation}
    r_{ij} =  \min \left\lbrace \left( \frac{k_{\mathrm{o}, i} - k_{\mathrm{o}, j}}{\sigma_k} \right)^2, \alpha \right\rbrace + \beta \delta_{\ell_{i} \ell_{j}}.
\end{equation}
Finally, we define $L_{\mathrm{M}}$:
\begin{equation}
    L_{\mathrm{M}}(\mathrm{M}) = 10 \sum_i \left( \sum_j \mathrm{M}_{ij} -1 \right)^2
\end{equation}
to regulate the fit by maintaining the original normalization of the window. We minimize the loss function $L$ with \texttt{jax}\footnote{\url{https://github.com/google/jax}.} and \texttt{optax}\footnote{\url{https://github.com/google-deepmind/optax}.} Python packages. Since we want to suppress elements of the window matrix above $\kt \sim 0.2 \; h/\mathrm{Mpc}$, we initialize $\vmt$ to zero at $\kt < 0.2 \; h/\mathrm{Mpc}$ and to the $\ko\ \sim 0.1 \; h/\Mpc$ row of $\mathrm{W}$ for $\kt\ > 0.2 \; h/\Mpc$ (we could choose any other row). We then initialize $\vmo$ to the last column of each $\ell_{\mathrm{t}} = \ell_{\mathrm{o}}$ diagonal block of W over the last element of the corresponding $\vmt$ multipole, so that the initial window $\mathrm{W}^{\prime}$ is set to zero at the maximum $\kt$.

In order to prevent the optimization procedure from being affected by the noise in the EZmocks covariance, we use a Gaussian analytical covariance matrix to tune M, $\vmo$ and $\vmt$. The analytical covariance matrix was generated with \texttt{thecov}~\cite{KP4s8-Alves}.\footnote{\url{https://github.com/cosmodesi/thecov}.} Note that this covariance matrix does not include the $\theta$-cut effect, contrary to the EZmocks covariance matrix. However here it is only used to fit rotation parameters. We then use the EZmocks covariance matrix transformed through M, fitted on the analytical covariance matrix, as $\mathrm{C}^{\prime}$.

\begin{figure}
\centering
\includegraphics[scale=0.8]{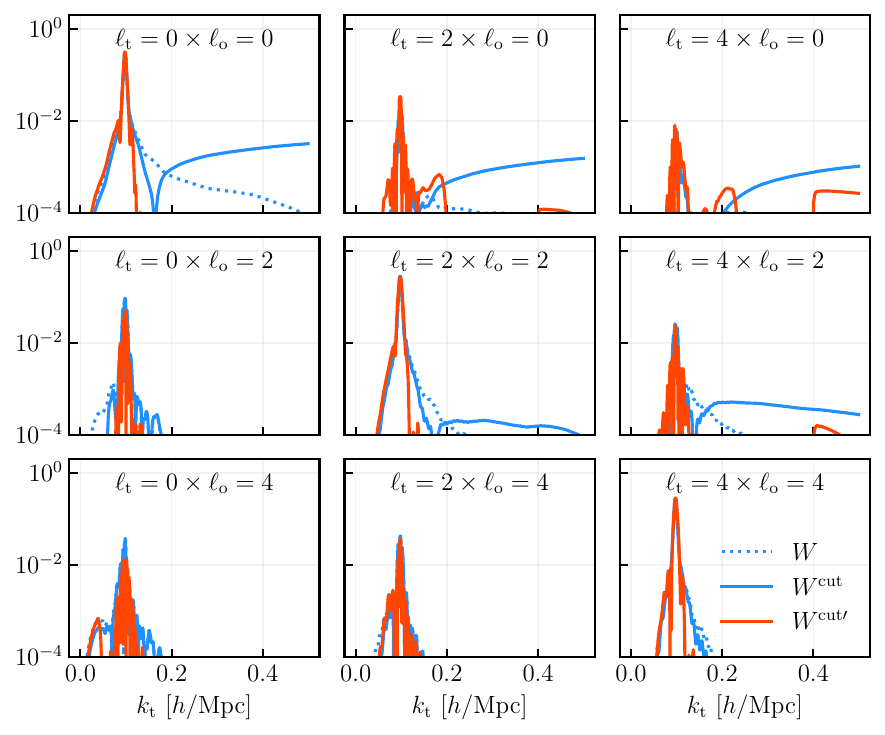}
\caption{Absolute value of the transformed $\theta$-cut window matrix (red) at \ko\ $= 0.1 \; h/\Mpc$, normalized by \kt, compared with that of the original uncut (dotted blue) and $\theta$-cut (blue) window matrices. The transformation suppresses the diverging tails at \kt\ $> 0.2 \; h/\Mpc$ from the $\theta$-cut window matrix.}
\label{fig:sculptedwindow}
\end{figure}

Figure~\ref{fig:sculptedwindow} shows the transformed $\theta$-cut window matrix $\mathrm{W}^{\cut \prime}$, together with the original uncut and $\theta$-cut windows W, $\mathrm{W}^{\cut}$, for a fixed $\ko = 0.1 \; h/\Mpc$. Note that figure~\ref{fig:sculptedwindow} shows the absolute value of the window divided by $\kt$ to compensate visually for the log-space $\kt$-spacing of the window. The original uncut window converges to 0 at high $\kt$, but the $\theta$-cut window has diverging tails, especially for the input monopole ($\ell_{\mathrm{o}} = 0$). The transformation almost completely suppresses these tails for $\mathrm{W}^{\cut \prime}$, although some residuals are left at $\kt > 0.4 \; h/\Mpc$. Note however that the relative scale of the tail in the $(\ell_{\mathrm{t}}, \ell_{\mathrm{o}}) = (4, 0)$ term, where it is the highest, is at least one order of magnitude below the magnitude of the peak at $\kt\ = 0.1 \; h/\Mpc$, which is itself around two orders of magnitude below the monopole ($(\ell_{\mathrm{t}}, \ell_{\mathrm{o}}) = (0, 0)$) peak. 

\begin{figure}
\centering
\includegraphics[scale=0.8]{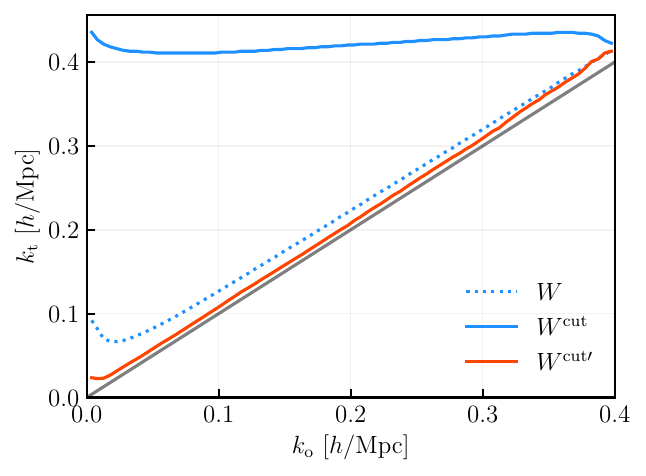}
\caption{For each \ko, value of \kt\ such that $\sum_{k=0}^{k_{\mathrm{t}} } |\mathrm{W}(k_{\mathrm{o}}, k)| < 0.95 \sum_{k} |\mathrm{W}(k_{\mathrm{o}}, k)|$, where $\mathrm{W}(k_{\mathrm{o}}, k)$ is the power spectrum window matrix, in three cases. The black line delineates the diagonal \kt\ = \ko. 95\% of the absolute weight of the transformed $\theta$-cut window $\mathrm{W}^{\cut \prime}$ (red) is comprised within \kt\ $\lesssim 0.2 \; h/\Mpc$ as for the original uncut matrix, $\mathrm{W}$ (dotted blue) whereas it is spread out up to \kt\ $\sim 0.4 \; h/\Mpc$ in the case of the non-transformed $\theta$-cut
matrix $\mathrm{W}^{\cut}$ (solid blue).}
\label{fig:sculptedwindow_compactness}
\end{figure}

Figure~\ref{fig:sculptedwindow_compactness} assesses the compactness of the transformed window matrix. The vertical axis corresponds to the minimum value of $\kt$ up to which we need to integrate the window matrix in order to capture at least 95\% of the total absolute weight of the window, for each $\ko$. We see that for all $\ko$, the minimum value of $\kt$ required to get 95\% of the window weight is close to $\ko$, similarly to the original uncut window matrix $\mathrm{W}$, whereas the  non-transformed $\theta$-cut window matrix weights are spread out on all $\kt$ modes up to $\kt\ \sim 0.4 \; h/\Mpc$. In particular, for all $\ko$ values below $0.2 \; h/\Mpc$, which is the scale cut used in the fits, 95\% of the window absolute weight is comprised within $\kt \lesssim 0.2 \; h/\Mpc$.

\begin{figure}
\centering
\includegraphics[scale=0.78]{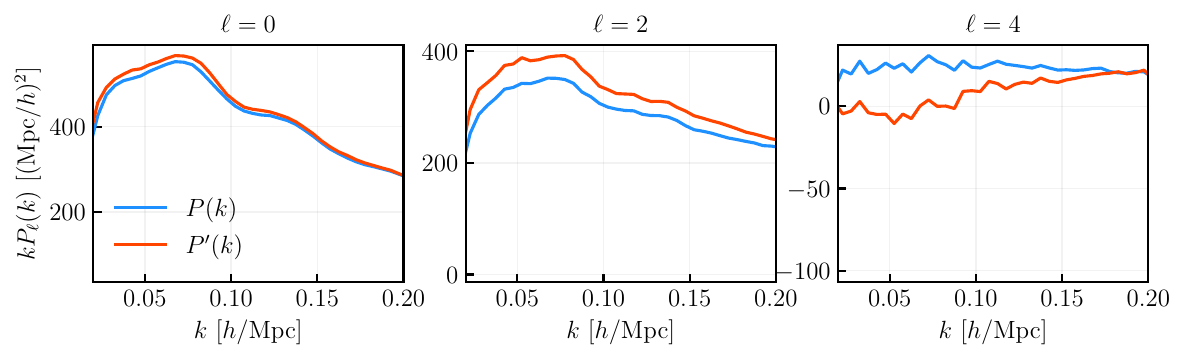}
\caption{Mutipoles ($\ell$ = 0, 2, 4) of the average $\theta$-cut power spectrum of the ELG mocks before (blue) and after (red) transformation of the window matrix to limit the numbers of high theory $k$ modes outside of the fitting range of full-shape fits. Here no $s\vmo$ term is added to the data vector, only M$\vd$ is shown.}
\label{fig:sculptedwindow_transformedpk}
\end{figure}

\begin{figure}
\centering
\includegraphics[scale=0.78]{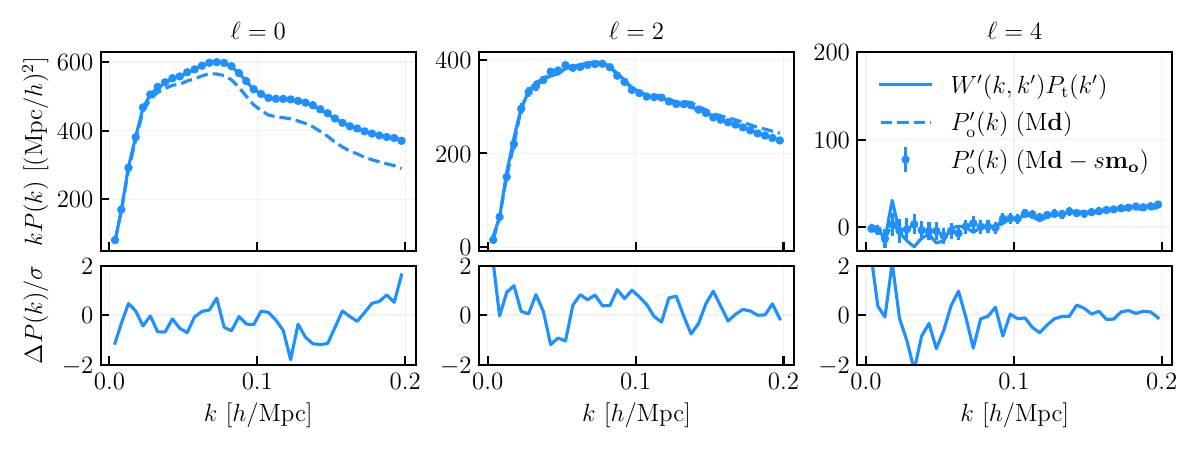}
\caption{Same as figure~\ref{fig:window_validation_complete_nocut} with $\theta$-cut applied and after transformation of the window matrix to limit the numbers of high theory $k$ modes outside of the full-shape fitting range.}
\label{fig:window_validation_complete_thetacut_sculptwindow}
\end{figure}

Figure~\ref{fig:sculptedwindow_transformedpk} shows how the transformation affects the power spectrum multipoles. The shape of the transformed power spectrum $\mathrm{M} \vd$ is only slightly modified compared to the original data vector $\vd$. Figure~\ref{fig:window_validation_complete_thetacut_sculptwindow} shows the transformed window times the cubic power spectrum multipoles together with the transformed cut-sky power spectrum multipoles. The $\mathrm{M}\vd$ term is shown as a dashed line, and $\mathrm{M}\vd - s\vmo$ where $s$ is fitted to minimize residuals is shown as dots. The best fit values for $s$ multipoles components are $s_{\ell}$ = -450, 15, -4 $(\mathrm{Mpc}/h)^3$ for $\ell$ = 0, 2, 4, respectively. The prediction for the observed power spectrum from the transformed window times the cubic power spectrum multipoles is still in good agreement with the transformed cut-sky power spectrum multipoles.

\begin{figure}
\centering
\includegraphics[scale=0.62]{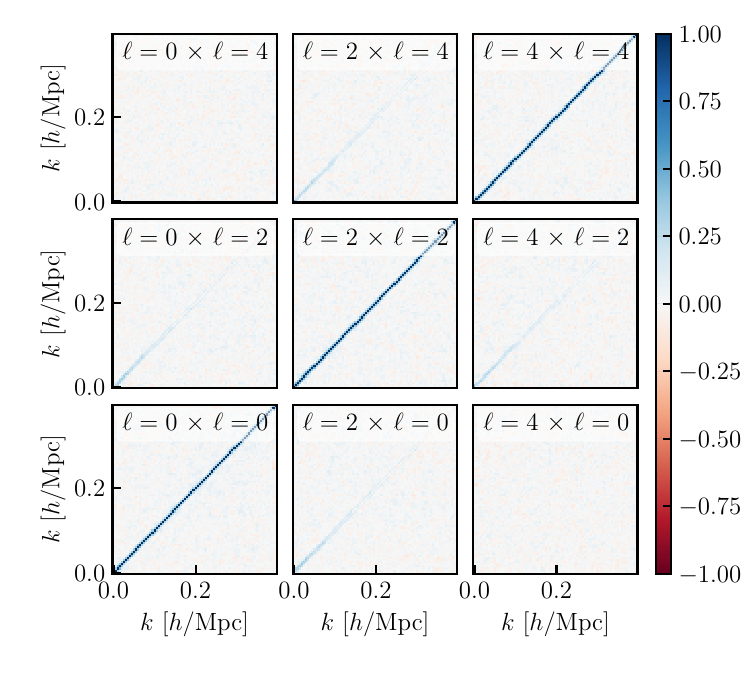}
\includegraphics[scale=0.62]{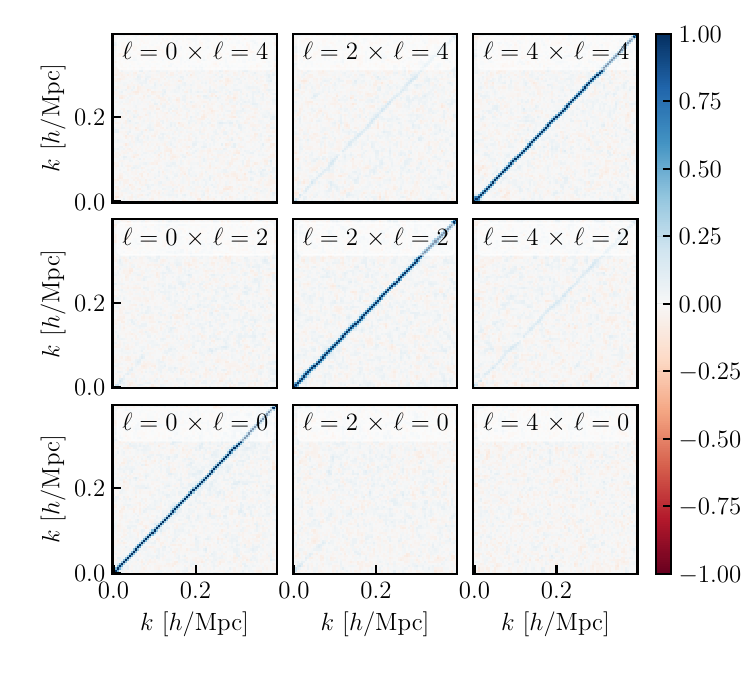}
\caption{\textit{Left:} $\theta$-cut power spectrum correlation matrix $\mathrm{R^{\cut}}$. \textit{Right:} transformed $\theta$-cut power spectrum correlation matrix $\mathrm{R^{\cut \prime}}$.}
\label{fig:sculptedcorrmatrix}
\end{figure}

Finally, figure~\ref{fig:sculptedcorrmatrix} shows that the covariance matrix is still close to diagonal after transformation, and that the off-diagonal multipole components of the transformed covariance matrix are slightly suppressed compared to that of the pre-transformation covariance matrix.

\subsection{Full-shape fits}
\begin{figure}
\centering
\includegraphics[scale=0.8]{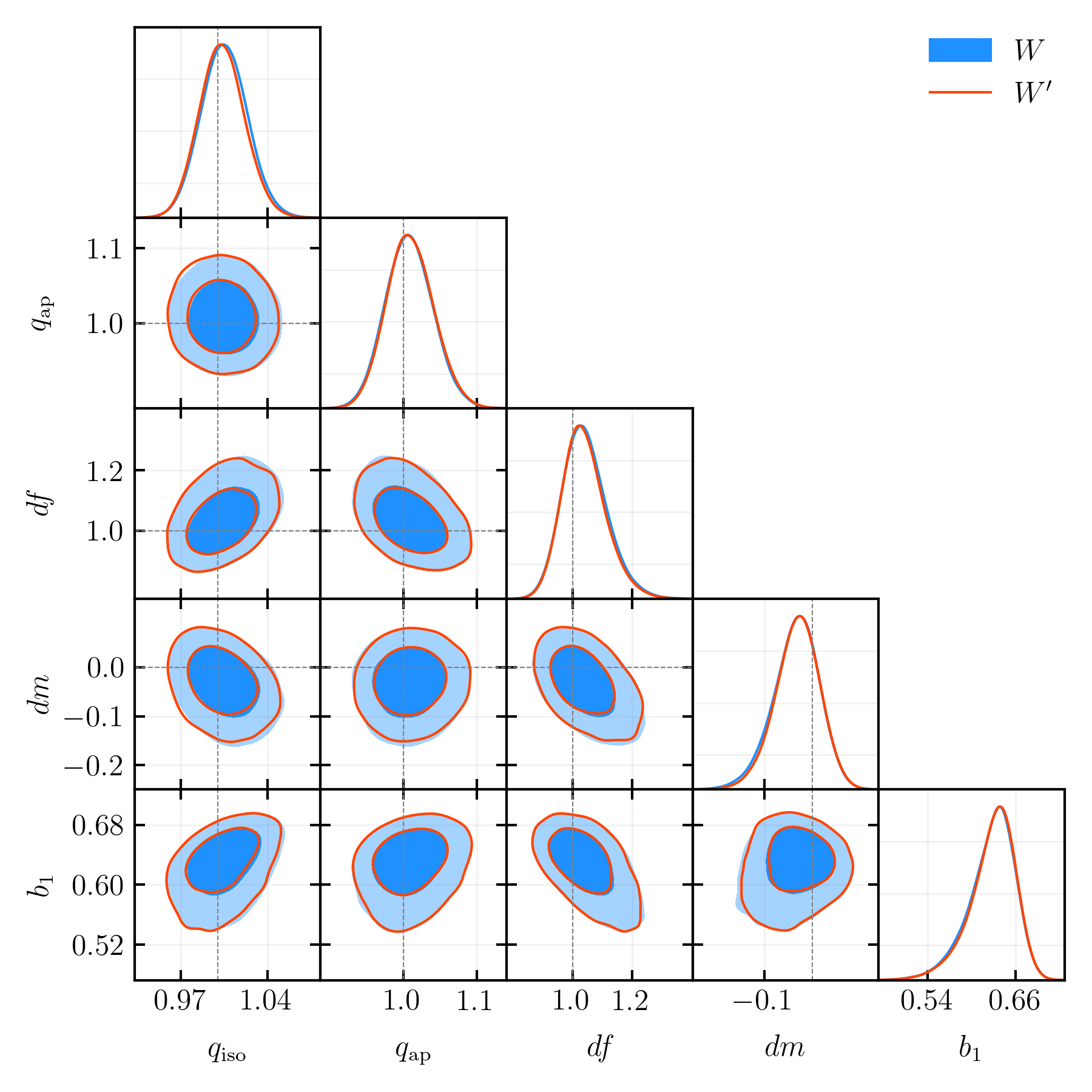}
\caption{
Marginalized posteriors on cosmological parameters $q_{\mathrm{iso}}$, $q_{\mathrm{ap}}$, $df$, $dm$ and linear bias $b_1$, obtained from sampling the average power spectrum multipoles ($\ell = 0, 2, 4$) of the 25 complete AbacusSummit cut-sky mocks with $\theta$-cut applied. Blue (resp. orange) contours use the power spectrum model with the original (resp. transformed) $\theta$-cut window matrix.
}
\label{fig:corner_plot_sculptedwindow_priors}
\end{figure}

We perform a full-shape fit with the new likelihood built from $\vd^{\prime}$, $\mathrm{W}^{\prime}$ and $\mathrm{C}^{\prime}$, in the same conditions as in section~\ref{sec:full_shape_fits}. Note that although the scale cuts ($k \in [0.02, 0.2] \; h/\Mpc$) are the same as in section~\ref{sec:full_shape_fits}, they are not equivalent in the rotated space. Hence the $\chi^2$ in equation~\eqref{eq:rotated_chi2} is not exactly the same as in section~\ref{sec:full_shape_fits}, so we might get different posteriors. Since the transformed window $\mathrm{M}\mathrm{W}$ is more diagonal that the original $\theta$-cut window, however, we expect these scale cuts to be better defined in the rotated space.
The additional parameters $s_{\ell}$ are left free in the fits. We impose Gaussian priors on $s_{\ell}$ with mean 0 and standard deviation equal to the values of $s_{\ell}$ found from a best fit where stochastic terms $s_{n, 0}, s_{n, 2}, s_{n, 4}$ are fixed to 0: -321, -8, -6 $(\mathrm{Mpc}/h)^3$ for $\ell$ = 0, 2, 4, respectively. We show in figure~\ref{fig:corner_plot_sculptedwindow_priors} that with these priors on $s_{\ell}$, the transformation does not increase posterior errors and the cosmological constraints are similar to before the transformation.

\begin{figure}
\centering
\includegraphics[scale=0.78]{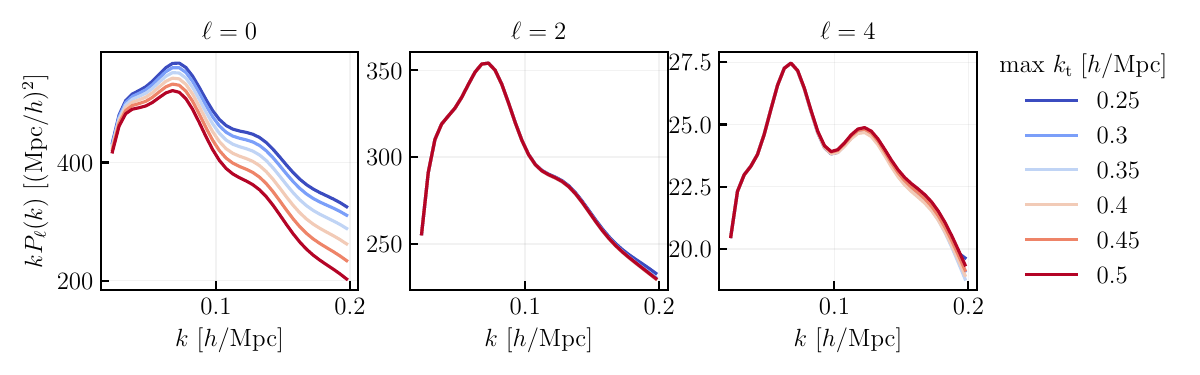}
\includegraphics[scale=0.78]{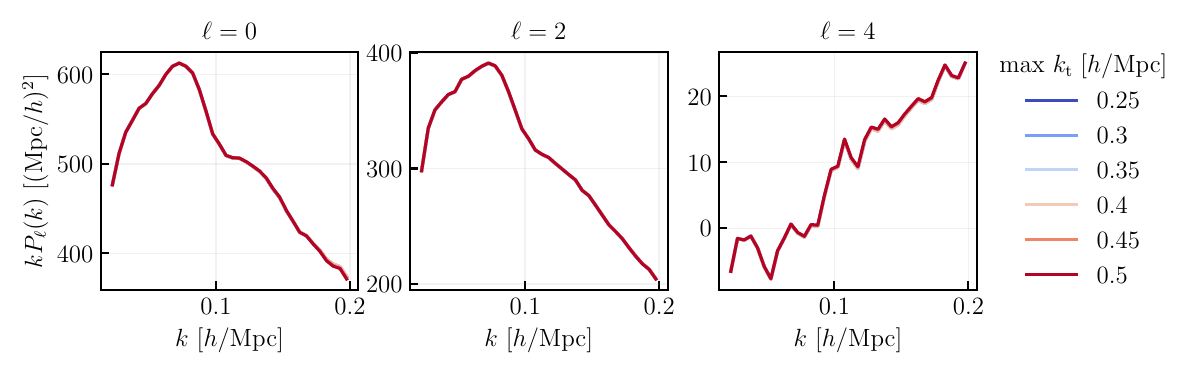}
\caption{Theoretical predictions for the observed power spectrum obtained by multiplying the window matrix W by the theoretical power spectrum (evaluated at the maximum posterior values of the parameters). Each line corresponds to a different cut in $\kt$: the legend on the right shows the values of the maximum $\kt$ used in the window. \textit{Top:} before applying the transformation M. \textit{Bottom:} after applying the transformation M.
}
\label{fig:theory_rotation_varyingktmax}
\end{figure}

Figure~\ref{fig:theory_rotation_varyingktmax} shows the cut-sky power spectrum multipoles prediction obtained by multiplying the theoretical power spectrum with the window matrix truncated at different values of $\kt$. The top panel shows the power spectrum multipoles with the window before the rotation, where we see that the multipoles prediction varies with the $\kt$-cut, whereas in the bottom panel, where the transformation $\mathrm{M}$ was applied, the power spectrum multipoles remain virtually unchanged when the maximum $\kt$ varies.

\section{Conclusion}
We have presented a method to mitigate the effect of fiber assignment incompleteness on two-point measurements, by cutting out small angular separations from two-point estimators. We present truncated two-point estimators implementing the cut in configuration and Fourier space, and we derive the corresponding modified windows allowing to model this cut.  We apply our method to high-accuracy N-body mocks designed to reproduce the clustering of the DESI DR1 ELG sample in the redshift bin $1.1 < z < 1.6$. We find that the truncation of small angular scales from two-point statistics allows us to recover unbiased cosmological constraints from mocks reproducing the fiber assignment incompleteness of DESI DR1, with respect to mocks with no fiber assignment incompleteness. The small size of the cut ($\theta < 0.05^{\circ}$) ensures that negligible statistical power is lost while removing all galaxy pairs affected by small-scale fiber assignment incompleteness.

Although the unbiased constraints obtained when cutting out small angular separations from the power spectrum suggests that any non-zero term at high-$\kt$ in the window matrix should be absorbed by stochastic parameters varied in the fit, we propose a method to remove such high-$\kt$ term arising from the truncation of small scales in the power spectrum window. Our approach consists in applying some rotation to the data vector, window matrix and covariance matrix, designed to suppress large-$\kt$ terms from the transformed window while leaving the likelihood unchanged. We define a loss function tailored to this end and fit the coefficients of the transformation matrix to minimize that loss function. We find that the transformation successfully suppresses high-$\kt$ terms while only slightly modifying the power spectrum. This transformation ensures that the window matrix, even when truncated at $\kt < 0.2 h/\mathrm{Mpc}$, correctly models the relationship between the underlying theory power spectrum and the observed power spectrum with survey geometry, imaging masks, and fiber assignment incompleteness corrections.

\section*{Data availability}
All the material needed to reproduce the figures of this publication is available at this site: \url{https://doi.org/10.5281/zenodo.11452140}.

\appendix

\section{Numerical computation of the contribution of small-\texorpdfstring{$\theta$}{theta} pairs}\label{sec:thetacut_powerspectrum_estimator_numerical_details}
\begin{figure}
\begin{center}
\includegraphics[scale=0.78]
{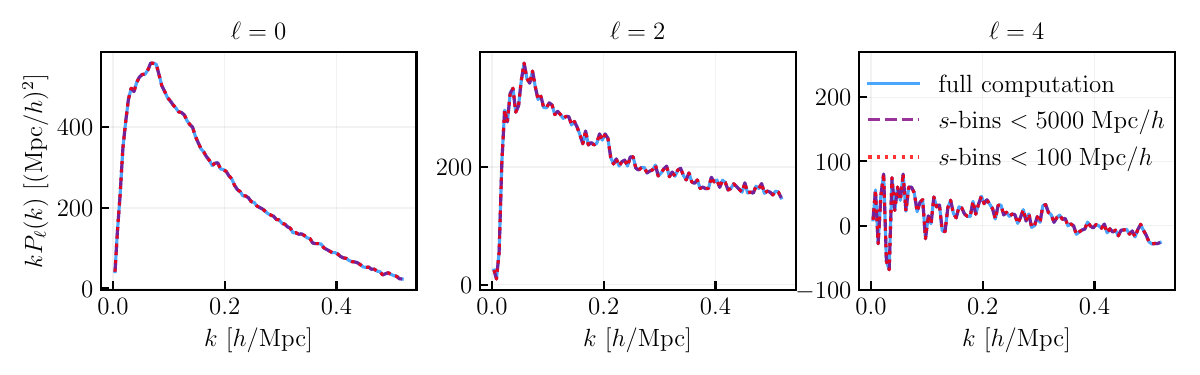}
\end{center}
\caption{Power spectrum multipoles computed on one AbacusSummit cut-sky ELG SGC mock, with 3 methods to remove low-$\theta$ pairs: summation over Bessel functions, $s$-binning with $s < 5000 \; \Mpc/h$ and $s$-binning restricting to $s < 100 \; \Mpc/h$.}
\label{fig:power_thetacut_computation_sbinning}
\end{figure}

The straightforward implementation of equation~\eqref{eq:power_direct} sums spherical Bessel functions (times an appropriate Legendre polynomial) for each pair of galaxies $i, \; j$ within $\theta < \Lambda_{\theta}$ and each mode $k$. With this implementation, the estimation of all pair counts takes more than one hour for one mock with randoms about $150$ times the size of the data on one Perlmutter node. To reduce computation time, we can rather bin pairs in separation $s$ (bin width $ds = 0.1 \; \Mpc/h$), and make the Bessel integral outside of the pair counts. Then the direct sums take approximately $80$ seconds, while the computation of the standard mesh power spectrum estimator~\eqref{eq:standard_power_estimator} typically takes about $190$ seconds. Besides, when $s$ increases, the total volume of the bin grows like $s^{2} ds$ while the volume of pairs with $\theta < \Lambda_{\theta}$ scales with $ds$, so the fractional contribution of these pairs decreases. As a result we can limit ourselves to separations $s < s_{\mathrm{max}}$. Restricting to $s_{\mathrm{max}} = 100 \; \mathrm{Mpc}/h$ helps us decrease the computing time for these direct sums to approximately $50$ seconds, with minimal changes to power spectrum estimates, as shown in figure~\ref{fig:power_thetacut_computation_sbinning}.

\section{Odd terms of the wide-angle expansion}\label{sec:wide_angle_odd}

In this section, we demonstrate equation~\eqref{eq:window_matrix_wa}. As noted by~\cite{reimberg_redshift-space_2016}, $\xi_{\ell}^{(n)}(s)$ and $P_{\ell}^{(n)}(k)$ coefficients in the wide-angle expansion are non-vanishing only if $\ell$, $n$ are either both odd or both even. Here we are computing the wide-angle expansion up to first order ($n = 1$), so let us compute odd correlation function and power spectrum poles. First, let us note that, by symmetry, we must have $\xi(\vx_{1},\vx_{2}) = \aver{\delta(\vx_{1})\delta(\vx_{2})} = \aver{\delta(\vx_{2})\delta(\vx_{1})} = \xi(\vx_{2},\vx_{1})$, such that, starting from equation~\eqref{eq:corr_wa}:
\begin{equation}
\xi(\vx_{1} = \vr,\vx_{2} = \vr + \vs) = \sum_{n,\ell} \left( \frac{s}{r} \right)^{n} \xi_{\ell}^{(n)}(s) \Leg{\ell}(\hat{\vr} \cdot \hat{\vs})
= \sum_{n,\ell} \left( \frac{s}{\abs{r + s}} \right)^{n} \xi_{\ell}^{(n)}(s) \Leg{\ell}(- \widehat{\vr + \vs} \cdot \hat{\vs}).
\end{equation}
At zeroth order in $s/r$, we have:
\begin{equation}
\sum_{\ell} \xi_{\ell}^{(0)}(s) \Leg{\ell}(\hat{\vr} \cdot \hat{\vs})
= \sum_{\ell} \xi_{\ell}^{(0)}(s) \Leg{\ell}(- \hat{\vr}\cdot \hat{\vs})
\end{equation}
so for $\ell$ odd, $\xi_{\ell}^{(0)}(s) = -\xi_{\ell}^{(0)}(s) = 0$. Now, at first order in $s/r$, $\widehat{\vr + \vs} \cdot \hat{\vs} = \mu + s/r \left(1 - \mu^{2}\right)$ with $\mu = \hat{\vr} \cdot \hat{\vs}$. Hence, expanding $\Leg{\ell}(\widehat{\vr + \vs} \cdot \hat{\vs})$ in Taylor series around $\mu$, and discarding second order terms in $s/r$, we get:
\begin{equation}
\sum_{\ell} \frac{s}{r} \xi_{\ell}^{(1)}(s) \Leg{\ell}(\mu)
= \sum_{\ell} \frac{s}{r} \xi_{\ell}^{(0)}(s) (-1)^{\ell} \Leg{\ell}^{\prime}(\mu) \left(1 - \mu^{2}\right) + \sum_{\ell} \frac{s}{r} \xi_{\ell}^{(1)}(s) \Leg{\ell}(- \mu).
\label{eq:legendre_taylor}
\end{equation}
By differentiating the Legendre polynomials' generating function:
\begin{equation}
g(\mu, t) = \frac{1}{\sqrt{1 - 2\mu t + t^2}}=\sum_{\ell=0}^{\infty}\Leg{\ell}(\mu)t^{\ell}
\end{equation}
with respect to $\mu$, we get that:
\begin{equation}
    \Leg{\ell}(\mu) + 2 \mu \Leg{\ell}^{\prime}(\mu) = \Leg{\ell + 1}^{\prime}(\mu) + \Leg{\ell - 1}^{\prime}(\mu).
\label{eq:legendre_deriv}
\end{equation}
Besides, Bonnet recurrence relation:
\begin{equation}
    (\ell + 1) \Leg{\ell+1}(\mu) = (2\ell + 1)\mu\Leg{\ell}(\mu) - \ell \Leg{\ell-1}(\mu)
\label{eq:bonnet}
\end{equation}
can be differentiated and combined with equation~\eqref{eq:legendre_deriv}, to obtain:
\begin{equation}
     \left(\mu^{2} - 1\right) \Leg{\ell}^{\prime}(\mu) = \ell \left[ \mu \Leg{\ell}(\mu) - \Leg{\ell-1}(\mu) \right].
\end{equation}
Using equation~\eqref{eq:bonnet} again, we get:
\begin{equation}
\left(\mu^{2} - 1\right) \Leg{\ell}^{\prime}(\mu) = \frac{\ell\left(\ell + 1\right)}{2\ell + 1} \left[\Leg{\ell+1}(\mu) - \Leg{\ell-1}(\mu)\right].
\end{equation}
Plugging this relation into equation~\eqref{eq:legendre_taylor}:
\begin{equation}
\sum_{\ell} \frac{s}{r} \xi_{\ell}^{(1)}(s) \left[\Leg{\ell}(\mu) - \Leg{\ell}(-\mu)\right] = \sum_{\ell} \frac{s}{r} \xi_{\ell}^{(0)}(s) (-1)^{\ell + 1} \frac{\ell\left(\ell + 1\right)}{2\ell + 1} \left[\Leg{\ell+1}(\mu) - \Leg{\ell-1}(\mu)\right].
\end{equation}
Equalizing terms of same $\Leg{\ell}(\mu)$, odd poles read:
\begin{equation}
\xi_{\ell}^{(1)}(s) = - \frac{\ell \left(\ell - 1\right)}{2 \left(2 \ell - 1\right)} \xi_{\ell-1}^{(0)}(s) + \frac{\left(\ell + 1\right) \left(\ell + 2\right)}{2 \left(2 \ell + 3\right)} \xi_{\ell+1}^{(0)}(s).
\label{eq:xi_first_order}
\end{equation}
Now let's derive the formula for power spectrum multipoles. Let us assume that:
\begin{equation}
P_{\ell}^{(1)}(k) = - \frac{\ell \left(\ell - 1\right)}{2 \left(2 \ell - 1\right)} A_{\ell}(k) + \frac{\left(\ell + 1\right) \left(\ell + 2\right)}{2 \left(2 \ell + 3\right)} B_{\ell}(k)
\end{equation}
and find $A_{\ell}(k)$ and $B_{\ell}(k)$. Starting from the definition of $P_{\ell}^{(n)}(k)$:
\begin{equation}
P_{\ell}^{(n)}(k) = 4 \pi i^{\ell} \int ds s^{2} (ks)^{n} \xi_{\ell}^{(n)}(s) j_{\ell}(ks)
\end{equation}
injecting~\eqref{eq:xi_first_order} (for $n=1$), and using that:
\begin{equation}
\xi_{\ell}^{(0)}(s) = \frac{(-i)^{\ell}}{2 \pi^{2}} \int dk k^{2} P_{\ell}^{(0)}(k) j_{\ell}(ks),
\end{equation}
we find:
\begin{equation}
A_{\ell}(k) = \frac{2i}{\pi} k \int ds s^{3} \int dk^{\prime} k^{\prime2} P_{\ell - 1}^{(0)}(k^{\prime}) j_{\ell - 1}(k^{\prime}s) j_{\ell}(ks).
\end{equation}
Following appendix F1 of~\cite{reimberg_redshift-space_2016}, let us define the following integrals:
\begin{equation}
\mathcal{I}_{p,q,n}(a,b) = \frac{2}{\pi} \int ds s^{2 + 2n + p + q} j_{p}(as) j_{q}(bs).
\end{equation}
We can thus rewrite:
\begin{equation}
A_{\ell}(k) = i k \int dk^{\prime} k^{\prime2} P_{\ell - 1}^{(0)}(k^{\prime}) \mathcal{I}_{\ell-1, \ell,  1-\ell}(k^{\prime},k).
\end{equation}
Using equations (F4) and (F10), we find:
\begin{align}
\mathcal{I}_{\ell-1, \ell, 1-\ell}(k^{\prime},k)
&= k^{\prime-\ell-1} \partial_{k^{\prime}} \left[ k^{\prime \ell + 1} \mathcal{I}_{\ell, \ell, -\ell}(k^{\prime},k) \right] \\
&= k^{\prime-\ell-1} \partial_{k^{\prime}} \left[ k^{\prime \ell + 1} \frac{\delta_{D}(k^{\prime} - k)}{k^{\prime2}} \right] \\
&= \left(\ell - 1\right) k^{\prime-3} \delta_{D}(k^{\prime} - k) + k^{\prime-2} \partial_{k^{\prime}} \delta_{D}(k^{\prime} - k),
\end{align}
hence:
\begin{equation}
A_{\ell}(k) = i \left[ \left(\ell - 1\right) P_{\ell - 1}^{(0)}(k) - k \partial_{k} P_{\ell - 1}^{(0)}(k) \right].
\end{equation}
We follow the same procedure for $B_{\ell}(k)$:
\begin{align}
B_{\ell}(k) &= - \frac{2i}{\pi} k \int ds s^{3} \int dk^{\prime} k^{\prime2} P_{\ell + 1}^{(0)}(k^{\prime}) j_{\ell + 1}(k^{\prime}s) j_{\ell}(ks) \\
&= - i k \int dk^{\prime} k^{\prime2} P_{\ell + 1}^{(0)}(k^{\prime}) \mathcal{I}_{\ell+1, \ell, -\ell}(k^{\prime},k),
\end{align}
and using equations (F6) and (F10) from~\cite{reimberg_redshift-space_2016}, we find:
\begin{align}
\mathcal{I}_{\ell+1, \ell, -\ell}(k^{\prime},k)
&= - k^{\prime \ell} \partial_{k^{\prime}} \left[ k^{\prime - \ell} \mathcal{I}_{\ell, \ell, -\ell}(k^{\prime},k) \right] \\
&= - k^{\prime \ell} \partial_{k^{\prime}} \left[ k^{\prime - \ell} \frac{\delta_{D}(k^{\prime} - k)}{k^{\prime2}} \right] \\
&= \left(\ell + 2\right) k^{\prime-3} \delta_{D}(k^{\prime} - k) + k^{\prime-2} \partial_{k^{\prime}} \delta_{D}(k^{\prime} - k).
\end{align}
Hence:
\begin{equation}
B_{\ell}(k) = - i \left[ \left(\ell + 2\right) P_{\ell + 1}^{(0)}(k) + k \partial_{k} P_{\ell + 1}^{(0)}(k) \right].
\end{equation}
In conclusion, we get the odd multipoles of the wide-angle expansion of $P_{\ell}(k)$, for $n=1$:
\begin{equation}
\begin{split}
P_{\ell}^{(1)}(k) = 
&- i \frac{\ell \left(\ell - 1\right)}{2 \left(2 \ell - 1\right)} \left[ \left(\ell - 1\right) P_{\ell - 1}^{(0)}(k) - k \partial_{k} P_{\ell - 1}^{(0)}(k) \right] \\
&- i \frac{\left(\ell + 1\right) \left(\ell + 2\right)}{2 \left(2 \ell + 3\right)} \left[ \left(\ell + 2\right) P_{\ell + 1}^{(0)}(k) + k \partial_{k} P_{\ell + 1}^{(0)}(k) \right].
\end{split}
\end{equation}

\section{Apodization}\label{sec:apodization}

An idea to make the window matrix more compact in theory space is to apply a smooth cut instead of the hard cut we have applied.

\begin{figure}
\begin{center}
\includegraphics[scale=0.8]{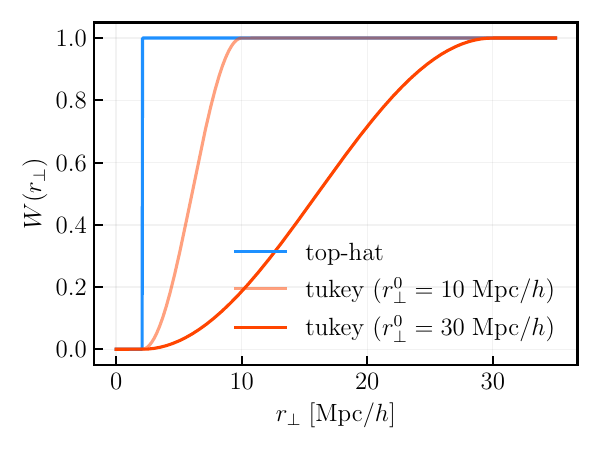}
\end{center}
\caption{Three windows to remove small transverse separation pairs: top-hat cut (blue), tukey cut with $r_{\perp}^0: 10 \; \mathrm{Mpc}/h$ (light red) and tukey cut with $r_{\perp}^0: 30 \; \mathrm{Mpc}/h$ (dark red).}
\label{fig:apodization}
\end{figure}

\begin{figure}
\begin{center}
\includegraphics[scale=0.8]{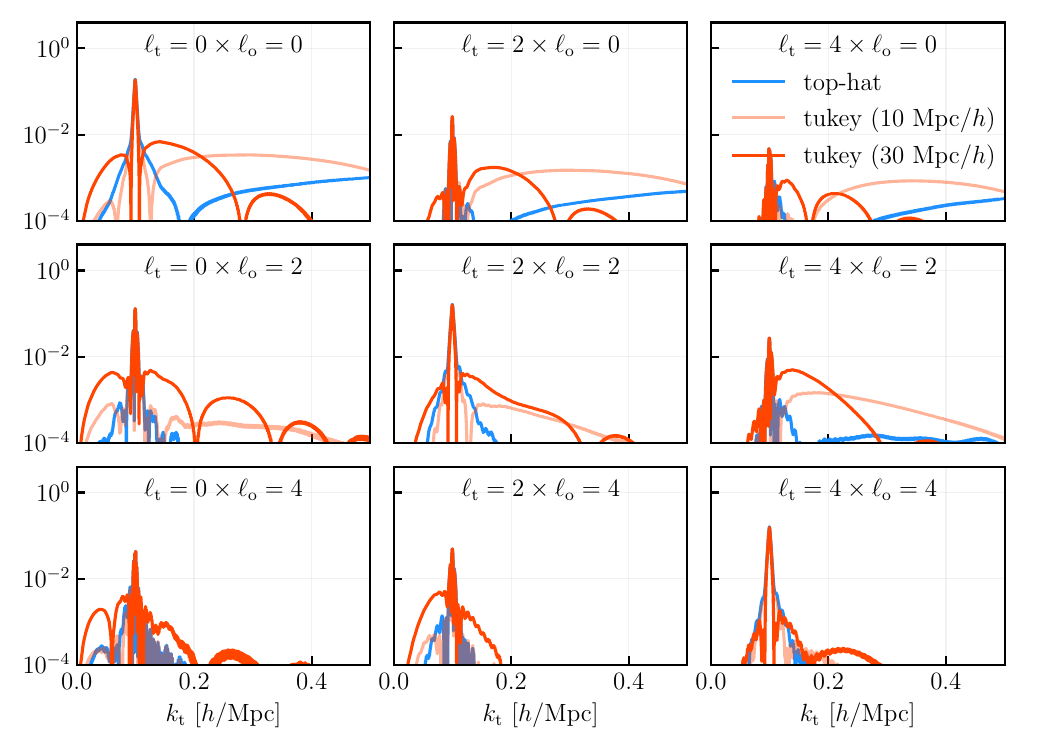}
\end{center}
\caption{Absolute value of the transformed window matrix (red) at \ko\ $= 0.1 \; h/\Mpc$, normalized by \kt with the three types of cuts shown in figure~\ref{fig:apodization}: top-hat cut (blue), tukey cut with $r_{\perp}^0: 10 \; \mathrm{Mpc}/h$ (light red) and tukey cut with $r_{\perp}^0: 30 \; \mathrm{Mpc}/h$ (dark red).}
\label{fig:wmatrix_apodized_rpcut}
\end{figure}

For simplicity in this section we consider a cut in separation transverse to the line-of-sight $r_{\perp}$ rather than a $\theta$-cut, as the two are equivalent in the small-angle approximation.
We estimate the $r_{\perp}$ value that corresponds to a $\Lambda_{\theta} = 0.05^{\circ}$ cut by computing the fraction of missing galaxy pairs shown in figure \ref{fig:dd_fiber_collisions} as a function of $r_{\perp}$.
Instead of a top-hat cut, i.e. a sharp truncation of all transverse separations below $\Lambda_{r_{\perp}}$, we explore a smoother cut using a tukey window, which is defined by:
\begin{equation}
    W_{\mathrm{tukey}}(r_{\perp}) =
    \begin{cases}
    0 & \text{if \;} 0 \leq r_{\perp} < \Lambda_{r_{\perp}} \\
    \frac{1}{2} \left[ 1 + \cos \left( \pi \frac{(r_{\perp}^0 - r_{\perp})}{(r_{\perp}^0 - \Lambda_{r_{\perp}})} \right) \right]  & \text{if \;} \Lambda_{r_{\perp}} \leq r_{\perp} < r_{\perp}^0 \\
    1 & \text{if \;} r_{\perp} \geq \ r_{\perp}^0
    \end{cases}
\end{equation}
with $\Lambda_{r_{\perp}} = 2.1 \; \mathrm{Mpc}/h$ and with two different values of $r_{\perp}^0$: $10 \; \mathrm{Mpc}/h$ and $30 \; \mathrm{Mpc}/h$. Figure~\ref{fig:apodization} shows the three windows (top-hat, tukey with $r_{\perp}^0: 10 \; \mathrm{Mpc}/h$, and tukey with $r_{\perp}^0: 30 \; \mathrm{Mpc}/h$) as a function of $r_{\perp}$. We then multiply the uncut configuration space window function with each of these windows, and compute the Fourier space window matrix. Figure~\ref{fig:wmatrix_apodized_rpcut} shows the resulting matrices at a fixed $\ko = 0.1 \; h/\mathrm{Mpc}$. We see that the small-scale cut induces oscillations in Fourier space that have an amplitude decreasing with $\kt$. The wider the cut is in configuration space, the higher the frequency of the oscillations are in Fourier space. We see that we need to apodize the cut up to scales of $\sim 30 \; \mathrm{Mpc}/h$ in order to get the Fourier space window to vanish above $\kt \sim 0.3-0.4 \; h/\mathrm{Mpc}$, which implies discarding information at scales much larger than the original cut we meant to do.

\section{Chained multipoles}\label{sec:chained_multipoles}

\begin{figure}
\begin{center}
\includegraphics[scale=0.8]{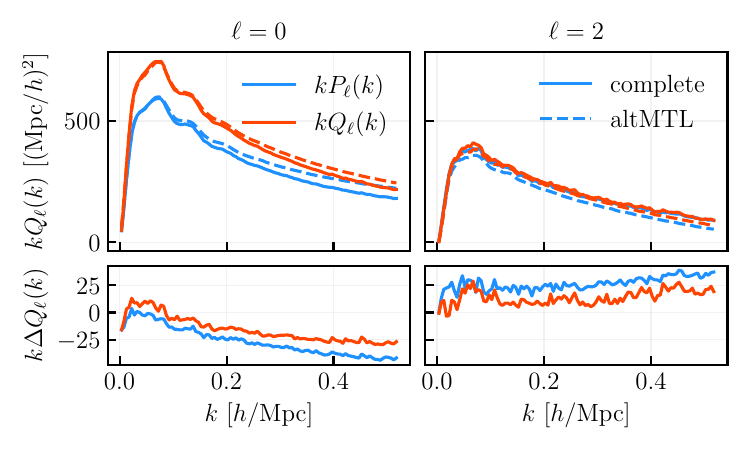}
\end{center}
\caption{\textit{Top}: chained (blue) and power spectrum (red) multipoles ($\ell$ = 0, 2) averaged over 25 DR1 ELG cut-sky mocks. Plain lines correspond to complete mocks and dashed lines to \altmtl\ (fiber-assigned) mocks. \textit{Bottom}: blue lines show the difference between chained multipoles in \altmtl\ and complete mocks, while red lines show the difference between power spectrum multipoles in \altmtl\ and complete mocks.}
\label{fig:chained_multipoles}
\end{figure}

Another idea to deal with spectroscopic systematics such as fiber assignment incompleteness effect, introduced in~\cite{zhao_completed_2021}, consists in subtracting transverse $\mu \sim 0$ modes with so-called chained multipoles. Assuming that fiber assignment incompleteness can be modelled as a power injection (or depletion) $X(k)$ at $\mu = 0$, the observed power spectrum can be written:
\begin{equation}
\hat{P}(k, \mu) = \hat{P}^{\mathrm{complete}}(k, \mu) + X(k) \delta_{D}(\mu)
\end{equation}
with $\hat{P}_{\ell}^{\mathrm{complete}}(k, \mu)$ the "clean" power spectrum (without fiber assignment incompleteness effect).
Taking the multipole of order $\ell$ (see~\cite{zhao_completed_2021}):
\begin{equation}
\hat{P}_{\ell}(k) = \hat{P}_{\ell}^{\mathrm{complete}}(k) + \frac{2 \ell + 1}{2} X(k) \Leg{\ell}(0),
\end{equation}
the next multipole order reads:
\begin{equation}
\hat{P}_{\ell + 2}(k) = \hat{P}_{\ell + 2}^{\mathrm{complete}}(k) + \frac{2 \ell + 5}{2} X(k) \Leg{\ell + 2}(0).
\end{equation}
Chained multipoles are thus defined as:
\begin{equation}
\hat{Q}_{\ell}(k) = \hat{P}_{\ell}(k) - \frac{(2\ell + 1) \Leg{\ell}(0)}{(2\ell + 5) \Leg{\ell + 2}(0)} \hat{P}_{\ell + 2}(k)
\end{equation}
such that the contamination $X(k)$ is removed. Figure~\ref{fig:chained_multipoles} show the comparison between the two first power spectrum multipoles and corresponding chained multipoles, averaged over the 25 ELG AbacusSummit cut-sky mocks. We see that the residuals between fiber-assigned and complete mocks is reduced but still significant in chained multipoles, suggesting that this procedure is not sufficient to remove fiber assignment incompleteness effect on the power spectrum.

\section{Author affiliations}
\label{sec:affiliations}

\noindent \hangindent=.5cm $^{1}${IRFU, CEA, Universit\'{e} Paris-Saclay, F-91191 Gif-sur-Yvette, France}

\noindent \hangindent=.5cm $^{2}${Lawrence Berkeley National Laboratory, 1 Cyclotron Road, Berkeley, CA 94720, USA}

\noindent \hangindent=.5cm $^{3}${Department of Physics, University of California, Berkeley, 366 LeConte Hall MC 7300, Berkeley, CA 94720-7300, USA}

\noindent \hangindent=.5cm $^{4}${University of California, Berkeley, 110 Sproul Hall \#5800 Berkeley, CA 94720, USA}

\noindent \hangindent=.5cm $^{5}${Dipartimento di Fisica ``Aldo Pontremoli'', Universit\`a degli Studi di Milano, Via Celoria 16, I-20133 Milano, Italy}

\noindent \hangindent=.5cm $^{6}${Center for Cosmology and AstroParticle Physics, The Ohio State University, 191 West Woodruff Avenue, Columbus, OH 43210, USA}

\noindent \hangindent=.5cm $^{7}${Department of Astronomy, The Ohio State University, 4055 McPherson Laboratory, 140 W 18th Avenue, Columbus, OH 43210, USA}

\noindent \hangindent=.5cm $^{8}${The Ohio State University, Columbus, 43210 OH, USA}

\noindent \hangindent=.5cm $^{9}${Physics Dept., Boston University, 590 Commonwealth Avenue, Boston, MA 02215, USA}

\noindent \hangindent=.5cm $^{10}${Department of Physics \& Astronomy, University College London, Gower Street, London, WC1E 6BT, UK}

\noindent \hangindent=.5cm $^{11}${Institute for Computational Cosmology, Department of Physics, Durham University, South Road, Durham DH1 3LE, UK}

\noindent \hangindent=.5cm $^{12}${Instituto de F\'{\i}sica, Universidad Nacional Aut\'{o}noma de M\'{e}xico,  Cd. de M\'{e}xico  C.P. 04510,  M\'{e}xico}

\noindent \hangindent=.5cm $^{13}${Department of Physics \& Astronomy and Pittsburgh Particle Physics, Astrophysics, and Cosmology Center (PITT PACC), University of Pittsburgh, 3941 O'Hara Street, Pittsburgh, PA 15260, USA}

\noindent \hangindent=.5cm $^{14}${Kavli Institute for Particle Astrophysics and Cosmology, Stanford University, Menlo Park, CA 94305, USA}

\noindent \hangindent=.5cm $^{15}${SLAC National Accelerator Laboratory, Menlo Park, CA 94305, USA}

\noindent \hangindent=.5cm $^{16}${Departamento de F\'isica, Universidad de los Andes, Cra. 1 No. 18A-10, Edificio Ip, CP 111711, Bogot\'a, Colombia}

\noindent \hangindent=.5cm $^{17}${Observatorio Astron\'omico, Universidad de los Andes, Cra. 1 No. 18A-10, Edificio H, CP 111711 Bogot\'a, Colombia}

\noindent \hangindent=.5cm $^{18}${Institut d'Estudis Espacials de Catalunya (IEEC), 08034 Barcelona, Spain}

\noindent \hangindent=.5cm $^{19}${Institute of Cosmology and Gravitation, University of Portsmouth, Dennis Sciama Building, Portsmouth, PO1 3FX, UK}

\noindent \hangindent=.5cm $^{20}${Institute of Space Sciences, ICE-CSIC, Campus UAB, Carrer de Can Magrans s/n, 08913 Bellaterra, Barcelona, Spain}

\noindent \hangindent=.5cm $^{21}${School of Mathematics and Physics, University of Queensland, 4072, Australia}

\noindent \hangindent=.5cm $^{22}${Department of Physics and Astronomy, University of California, Irvine, 92697, USA}

\noindent \hangindent=.5cm $^{23}${Department of Physics, Southern Methodist University, 3215 Daniel Avenue, Dallas, TX 75275, USA}

\noindent \hangindent=.5cm $^{24}${Sorbonne Universit\'{e}, CNRS/IN2P3, Laboratoire de Physique Nucl\'{e}aire et de Hautes Energies (LPNHE), FR-75005 Paris, France}

\noindent \hangindent=.5cm $^{25}${Departament de F\'{i}sica, Serra H\'{u}nter, Universitat Aut\`{o}noma de Barcelona, 08193 Bellaterra (Barcelona), Spain}

\noindent \hangindent=.5cm $^{26}${Institut de F\'{i}sica d’Altes Energies (IFAE), The Barcelona Institute of Science and Technology, Campus UAB, 08193 Bellaterra Barcelona, Spain}

\noindent \hangindent=.5cm $^{27}${NSF NOIRLab, 950 N. Cherry Ave., Tucson, AZ 85719, USA}

\noindent \hangindent=.5cm $^{28}${Instituci\'{o} Catalana de Recerca i Estudis Avan\c{c}ats, Passeig de Llu\'{\i}s Companys, 23, 08010 Barcelona, Spain}

\noindent \hangindent=.5cm $^{29}${Department of Physics and Astronomy, Siena College, 515 Loudon Road, Loudonville, NY 12211, USA}

\noindent \hangindent=.5cm $^{30}${Department of Physics \& Astronomy, University  of Wyoming, 1000 E. University, Dept.~3905, Laramie, WY 82071, USA}

\noindent \hangindent=.5cm $^{31}${Departamento de F\'{i}sica, Universidad de Guanajuato - DCI, C.P. 37150, Leon, Guanajuato, M\'{e}xico}

\noindent \hangindent=.5cm $^{32}${Instituto Avanzado de Cosmolog\'{\i}a A.~C., San Marcos 11 - Atenas 202. Magdalena Contreras, 10720. Ciudad de M\'{e}xico, M\'{e}xico}

\noindent \hangindent=.5cm $^{33}${Department of Physics and Astronomy, University of Waterloo, 200 University Ave W, Waterloo, ON N2L 3G1, Canada}

\noindent \hangindent=.5cm $^{34}${Perimeter Institute for Theoretical Physics, 31 Caroline St. North, Waterloo, ON N2L 2Y5, Canada}

\noindent \hangindent=.5cm $^{35}${Waterloo Centre for Astrophysics, University of Waterloo, 200 University Ave W, Waterloo, ON N2L 3G1, Canada}

\noindent \hangindent=.5cm $^{36}${Space Sciences Laboratory, University of California, Berkeley, 7 Gauss Way, Berkeley, CA  94720, USA}

\noindent \hangindent=.5cm $^{37}${Department of Physics and Astronomy, Sejong University, Seoul, 143-747, Korea}

\noindent \hangindent=.5cm $^{38}${CIEMAT, Avenida Complutense 40, E-28040 Madrid, Spain}

\noindent \hangindent=.5cm $^{39}${Department of Physics, University of Michigan, Ann Arbor, MI 48109, USA}

\noindent \hangindent=.5cm $^{40}${University of Michigan, Ann Arbor, MI 48109, USA}

\noindent \hangindent=.5cm $^{41}${Department of Physics \& Astronomy, Ohio University, Athens, OH 45701, USA}

\noindent \hangindent=.5cm $^{42}${National Astronomical Observatories, Chinese Academy of Sciences, A20 Datun Rd., Chaoyang District, Beijing, 100012, P.R. China}

\acknowledgments
This material is based upon work supported by the U.S. Department of Energy (DOE), Office of Science, Office of High-Energy Physics, under Contract No. DE–AC02–05CH11231, and by the National Energy Research Scientific Computing Center, a DOE Office of Science User Facility under the same contract. Additional support for DESI was provided by the U.S. National Science Foundation (NSF), Division of Astronomical Sciences under Contract No. AST-0950945 to the NSF National Optical-Infrared Astronomy Research Laboratory; the Science and Technology Facilities Council of the United Kingdom; the Gordon and Betty Moore Foundation; the Heising-Simons Foundation; the French Alternative Energies and Atomic Energy Commission (CEA); the National Council of Humanities, Science and Technology of Mexico (CONAHCYT); the Ministry of Science and Innovation of Spain (MICINN), and by the DESI Member Institutions: \url{https://www.desi.lbl.gov/collaborating-institutions}. Any opinions, findings, and conclusions or recommendations expressed in this material are those of the author(s) and do not necessarily reflect the views of the U. S. National Science Foundation, the U. S. Department of Energy, or any of the listed funding agencies.

The authors are honored to be permitted to conduct scientific research on Iolkam Du’ag (Kitt Peak), a mountain with particular significance to the Tohono O’odham Nation.

\bibliographystyle{JHEP}
\bibliography{biblio}{}

\end{document}